\documentclass[preprints,article,accept,pdftex,oneauthor]{mdpi} 

\firstpage{1} 
\pubvolume{1}
\issuenum{1}
\articlenumber{0}
\pubyear{2026}
\copyrightyear{2026}
\usepackage{subfig}
\usepackage[super]{nth}
\usepackage{fix-cm} 
\usepackage{siunitx}
\DeclareSIUnit\knot{kn}
\usepackage{arydshln} 
\usepackage{makecell}
\usepackage{tikz}
\usetikzlibrary{calc}
\usetikzlibrary{decorations.markings}
\usepackage{soul}

\def\UrlBreaks{\do\/\do-}

\Title{High-Resolution Mapping of Port Dynamics from Open-Access AIS Data in Tokyo Bay}

\Author{Moritz 
 H\"utten \orcidA{}}

\AuthorNames{Moritz H\"utten}

\address[1]{%
GRID Inc., 3-6-7 Kita-Aoyama, Minato-ku, Tokyo 107-0061, Japan; moritz.huetten@gridpredict.co.jp}

\abstract{Knowledge about vessel activity in port areas and around major industrial zones provides insights into economic trends, supports decision-making for shipping and port operators, and contributes to maritime safety. Vessel data from terrestrial receivers of the Automatic Identification System (AIS) have become increasingly openly available, and we demonstrate that such data can be used to infer port activities at high resolution and with precision comparable to official statistics. We analyze open-access AIS data from a three-month period in 2024 for Tokyo Bay, located in Japan's most densely populated urban region. Accounting for uneven data coverage, we reconstruct vessel activity in Tokyo Bay at $\sim\,$30~m resolution and identify 161 active berths across seven major port areas in the bay. During the analysis period, we find an average of $35\pm17_{\text{stat}}$ vessels moving within the bay at any given time, and $293\pm22_{\text{stat}}+65_{\text{syst}}-10_{\text{syst}}$ vessels entering or leaving the bay daily, with an average gross tonnage of $11{,}860^{+280}_{- \;\,50}$. These figures indicate an accelerating long-term trend toward fewer but larger vessels in Tokyo Bay’s commercial traffic. Furthermore, we find that in dense urban environments, radio shadows in vessel AIS data can reveal the precise locations of inherently passive receiver stations.}

\keyword{open data; Automatic Identification System (AIS); vessel stop detection; port performance; berth identification; radio shadowing} 

\addhighlights{yes}
\renewcommand{\addhighlights}{%
\vspace{3pt}\\
\textbf{What 
 are the main findings?}
\begin{itemize}[labelsep=2.5mm,topsep=-3pt]
\item The trend toward fewer and larger vessels entering Tokyo Bay is accelerating.
\item Passive AIS receiver stations in urban environments can be precisely localized through radio-shadow patterns.
\end{itemize}\vspace{3pt}
\textbf{What is the implication of the main findings?}
\begin{itemize}[labelsep=2.5mm,topsep=-3pt]
\item Open-access positioning data can accurately 
 map vessel traffic and berth activity in port areas and around major coastal industrial zones.
\end{itemize}
}

\begin{document}


\section{Introduction}
\label{sec:intro}

Tokyo Bay in Japan is one of the world's most crowded maritime zones \citep{Shimizu2020}. Around 20\% of Japan's total trade volume by weight and 30\% of its value---equivalent to 1\% of global trade value in 2024---are handled through the ports in Tokyo Bay, which at its narrowest point is only \SI{7}{\km} wide \citep{JCG2017,JapanFinance2015,UNCTAD2024,Jiang2025}. Despite its economic relevance, there is still little scientific literature analyzing maritime activity in this area. The few existing studies include basic explorations using data from the Automatic Identification System (AIS)  \citep{Takahashi2007,Tasseda2014,Shirai2016,Hyodo2017}, traffic control \citep{,Shimizu2020,Song2022}, collision avoidance \citep{TokyoWanAssociation2018,Terayama2020,Itoh2022,Kawashima2022,Kim2023}, environmental aspects \citep{Sakurai2021}, or berth occupancy \citep{Koizumi2015} in Tokyo Bay. The \citet{JCG2017} published a thorough study of traffic, turnover at ports of call, and hazardous areas within Tokyo Bay based on AIS data.

In recent years, AIS has proven to be the most valuable data source for live and historical studies of maritime activities \cite{Svanberg2019}. With respect to port operations, AIS data have been used to monitor port entries and exits \cite{Li2022}, differentiate handled vessel types~\citep{Rajabi2019} and movements within ports \citep{Lee2021,Feng2020,Martincic2020}, identify stopping points like berths or anchorages~\citep{Yan2022,Steenari2022,AsianDevelopmentBank2023,VanDerWielen2024,Wijaya2024,Iphar2024,Qiang2025}, estimate stopover durations \cite{Rao2025}, and distinguish between time spent waiting and actively berthing~\citep{Franzkeit2020,Ma2023}. Other studies focus on particular operations, such as container handling~\citep{Polimeni2024,Yasuda2024,Feng2024,Kim2025}, bulk cargo throughput \cite{Nakashima2024}, or bunkering services \cite{Watanabe2023}. Insights at sub-port resolution can also be obtained from open-access AIS data, as recently shown by \cite{Hadjipieris2025,Belcore2025}. Complementary to AIS data, satellite imagery, particularly using synthetic-aperture radar (SAR) \cite{ElDarymli2013}, has been widely used in maritime research and can detect vessels that do not actively transmit their positions \cite{Li2025}. While SAR-based vessel detection has also already been applied to Tokyo Bay \cite{Ouchi2013,Marino2015}, SAR may suffer, especially in coastal areas, from radar  shadowing \cite{Chaturvedi2019,Luo2023} and false-positive detections caused by confusion with other man-made objects, rocks in the surf \cite{Paolo2024}, or speckle noise \cite{Ying2024}. In addition, unlike AIS data, SAR imagery does not provide auxiliary information such as port of call or vessel identity.

In \citep{Huetten2025a}, we presented a thorough study of how open-access AIS data 
from terrestrial receivers can be effectively used to reconstruct maritime activity in coastal waters. Here, we extend this analysis to sub-port resolution ($\lesssim\,$\SI{30}{\meter}) using AIS data from Tokyo Bay, from the same provider, and for the same period of August to October 2024. Our high-resolution mapping highlights two technical aspects. First, the spatial coverage of openly available data from Tokyo Bay is highly uneven, and data availability continuously fades toward the open sea. This makes it challenging to correctly differentiate transiting vessels, i.e., vessels that enter or exit the region of interest (ROI), from moored vessels scattered throughout the ROI while waiting before or after port handling. Distinguishing absence from mooring adds an important aspect to existing approaches to identify vessel stops in port-size, spatially limited ROIs. Second, our AIS data reception is patchy and fluctuates throughout the bay, shadowed by tall buildings between vessels and receivers in the surrounding urban area. These shadows allow  the reconstruction of AIS receiver positions in such an environment. 
 Finally, we provide recent estimates, along with their uncertainties, of vessel activity in Tokyo Bay and its trend by comparison to data from the past.

This paper is structured as follows. \Cref{sec:data_acquisition} describes the analysis region and the AIS data used. In \Cref{sec:data_processing,sec:vessel_metrics,sec:methods_basestations}, we outline the data-processing methods specific to Tokyo Bay and how AIS receiver positions are inferred from vessel positioning data. In \Cref{sec:results_vessels}, we present the results on traffic patterns, average vessel counts, and the most frequently used berths in the bay, and \Cref{sec:results_basestation} presents the results of the receiver-position inference. We discuss our findings in \Cref{sec:discussion} and conclude this paper in \Cref{sec:summary}.


\section{Methods}
\label{sec:methods}

\subsection{Analysis Region and Data Acquisition}
\label{sec:data_acquisition}

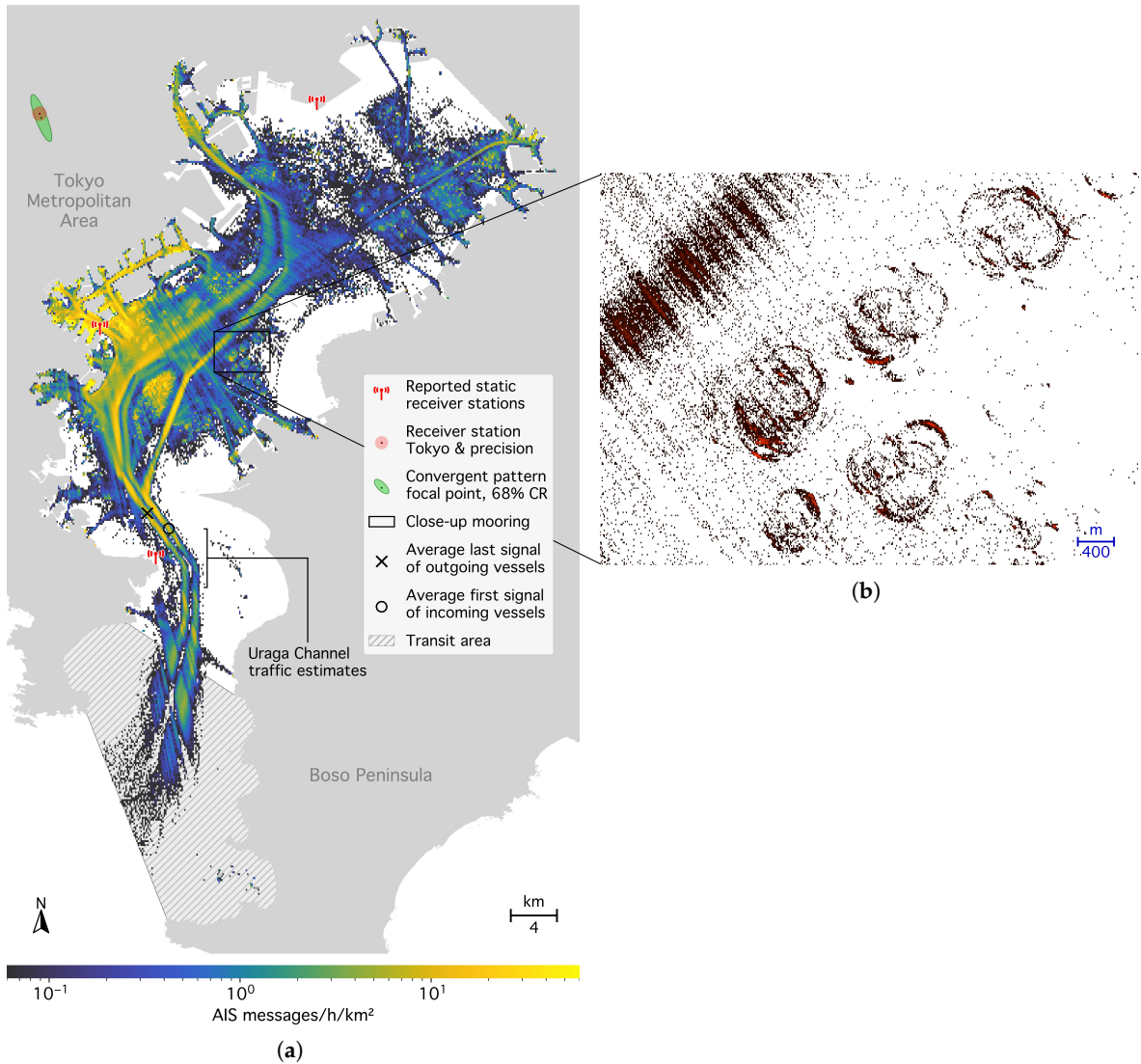
\begin{figure}[ht!] 
\captionsetup[subfloat]{margin=2pt}

	\centering
	\subfloat[  
 \label{fig:roi_data_map_tokyo}]{
        \tikz[remember picture]{
            \node[inner sep=0pt,anchor=south west] (A)
            {\includegraphics[width=0.5\textwidth,
                trim=0px 0px 0px 15px,clip]{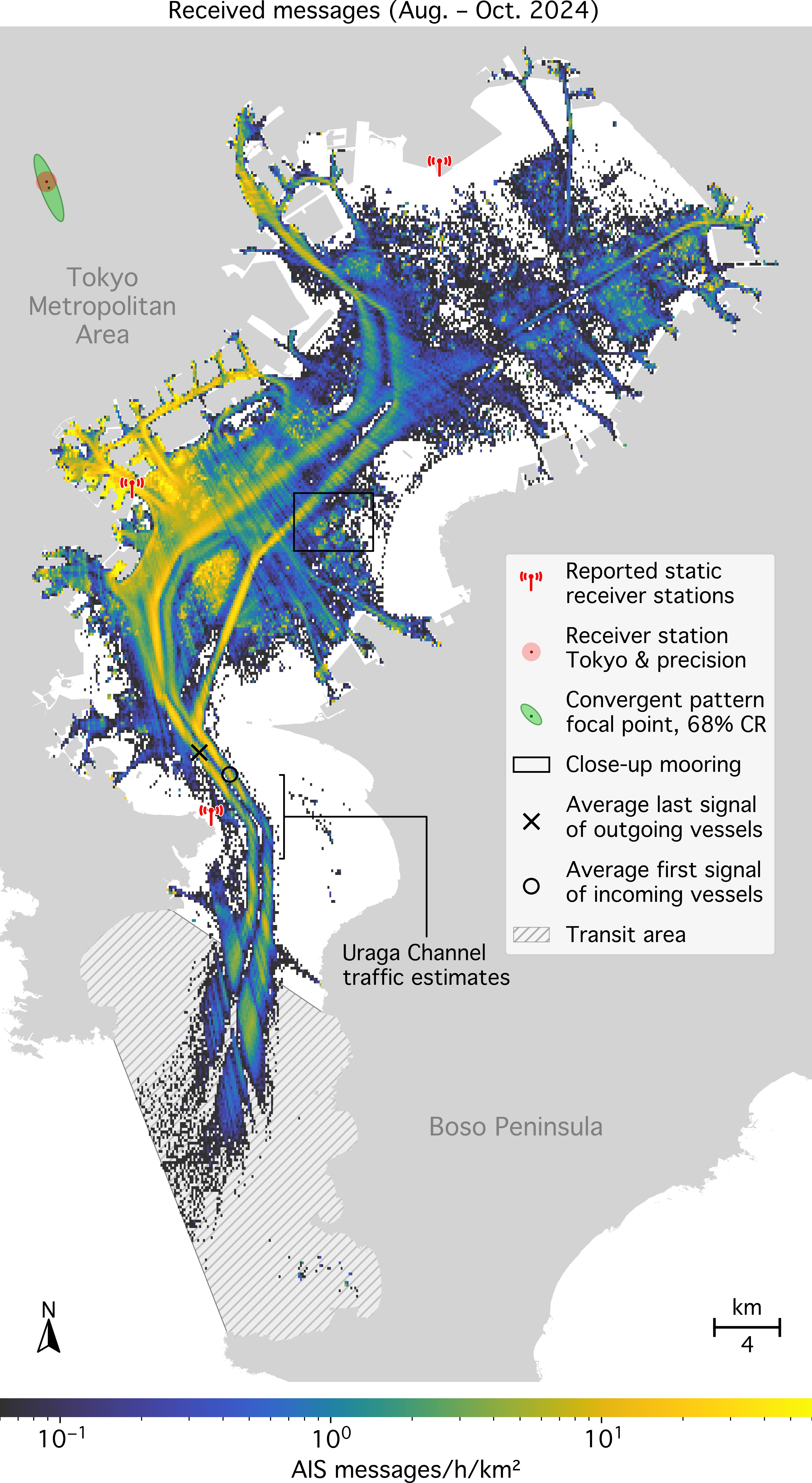}};
        }
    }
    \hfill
    		\raisebox{1.1\height}{\subfloat[      
		 \label{fig:roi_data_map_tokyo_cutout}
		]{
        \tikz[remember picture]{
            \node[inner sep=0pt,anchor=south west] (B)
            {\includegraphics[width=0.47\textwidth]{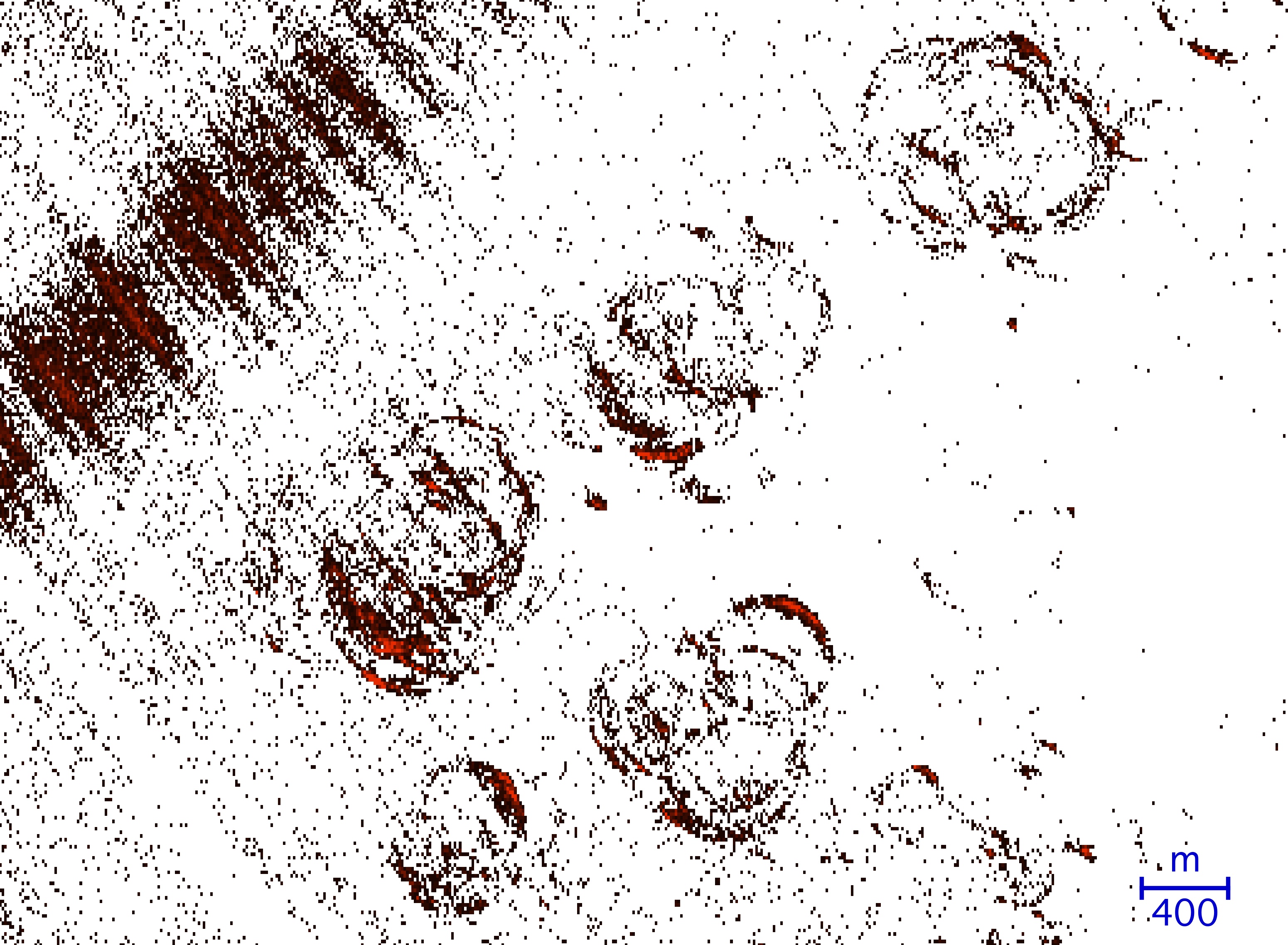}};
        }
    }}

	  \caption{Tokyo Bay region of interest (ROI) and positions of AIS-A messages received between 29 July and 27 October 2024. (\textbf{a}) Message frequency in the ROI. The panel covers longitudes $139.60^\circ\,\mathrm{E}$ to $140.15^\circ\,\mathrm{E}$ and latitudes $34.95^\circ\,\mathrm{N}$ to $35.70^\circ\,\mathrm{N}$. The gray-hatched area (lower part of the left panel) indicates the transit area of entering or leaving vessels. Additionally, three AIS receiver stations reported by the network are shown in red. The position of the Tokyo receiver with precision from AISHub \cite{AISHub2025} is shown as a red disk in the left panel, and the inferred position and the 68\% containment range (CR) as a green ellipse (Section$\,$\ref{sec:results_basestation}). Average positions where first contact occurs with entering (black circle) and last contact with leaving (black diagonal cross) vessels are also shown, along with the zone where the \citet{JCG2020} estimates traffic through the Uraga Channel.
(\textbf{b}) Close-up with message positions binned at higher resolution. Redder (brighter) colors indicate more messages per bin. The close-up illustrates radio shadows from signal occlusion by buildings {(
 Section~\ref{sec:methods_basestations})} and circular patterns from vessels swaying around their anchor chains while moored \citep{Wijaya2024,Visky2024,Koizumi2015}.
	  }
    \label{fig:roi_data_map_tokyo_both}

\begin{tikzpicture}[remember picture, overlay]

\def\yOffset{2.16} 

\begin{scope}[yshift=\yOffset cm]

\def\xStart{-4.89}
\def\xEnd{0.38}
\def\yStart{12.645}
\def\yEnd{10.01}

\def\yStarttwo{13.205}
\def\yEndtwo{15.362}

\def\gapStart{0.3875}
\def\gapEnd{0.877}

\pgfmathsetmacro{\xGapStart}{\xStart + \gapStart*(\xEnd-\xStart)}
\pgfmathsetmacro{\yGapStart}{\yStart + \gapStart*(\yEnd-\yStart)}
\pgfmathsetmacro{\xGapEnd}{\xStart + \gapEnd*(\xEnd-\xStart)}
\pgfmathsetmacro{\yGapEnd}{\yStart + \gapEnd*(\yEnd-\yStart)}

\draw[thin, black] (\xStart,\yStart) -- (\xGapStart,\yGapStart);
\draw[thin, black] (\xGapEnd,\yGapEnd) -- (\xEnd,\yEnd);
\draw[thin, black] (\xStart,\yStarttwo) -- (\xEnd,\yEndtwo);

\end{scope}
\end{tikzpicture}
    
\end{figure}

Tokyo Bay (Tokyo Wan
) is located on the coast of Honshu, the main island of Japan, encircled by the Tokyo Major Metropolitan Area in the west and the Boso Peninsula in the east, as illustrated in Figure \ref{fig:roi_data_map_tokyo_both}a. It extends about \SI{60}{\km} from north to south and is, on average, approximately \SI{15}{\km} wide in the east--west direction. Geologically, Tokyo Bay is divided into an inner bay north of Cape Kannon in the west and Cape Futtsu in the east, and an outer bay extending further south through the Uraga Channel (Uraga Suido, as shown in Figure~\ref{fig:roi_data_map_tokyo_both}a), where the seabed begins to deepen into the Pacific Ocean. According to Japanese Maritime Traffic Laws \citep{JCG2017}, the outer bay is demarcated from the open sea by a line between the Tsurugisaki Lighthouse in the west and the Susaki Lighthouse in the east. We use this boundary as the outer limit of our ROI, corresponding to the lower edge of the hatched transit area in Figure~\ref{fig:roi_data_map_tokyo_both}a. The area enclosed by this line and the shorelines from  \citet{Openstreetmap2024} 
 covers a water surface of \SI{1344}{\square\km}. The northern edge of the transit area in Figure~\ref{fig:roi_data_map_tokyo_both}a is defined by a line between the Ashikashima Lighthouse in the west and Myogane Cape in the east. We consider all activity north of this line as occurring within the bay. This choice results in traffic from or to Tateyama Bay (indicated by isolated messages at the bottom of Figure~\ref{fig:roi_data_map_tokyo_both}a) being counted as inbound or outbound from the ROI, but counts the Tokyo Bay Ferry between Kurihama and Kanaya as traffic within the bay. Several rivers discharge into Tokyo Bay, the largest being the Arakawa with its Sumidagawa branch, the Tamagawa, and the Edogawa. As these waterways are closed to most shipping, we exclude entries or exits (collectively referred to as ``transits'' hereafter) through their mouths.

 \vspace{-6pt}

Tokyo Bay contains several of Japan's largest ports. These include the ports of Tokyo, Kawasaki, and Yokohama, the three ports forming the Keihin (a blend of Tokyo and Yokohama) port area with a nearly continuous \SI{50}{\km}-long waterfront. Also in Tokyo Bay lie the ports of Funabashi, Chiba City, Kisarazu, and Yokosuka, the last hosting the largest overseas naval facility of the United States \citep{CNIC2025}. Together, these ports cover approximately \SI{560}{\square\km}~\citep{JCG2017}, which is more than one-third of the bay’s water area. 

 We maintained an uninterrupted subscription to the public AIS data stream from \citet{aisstream2025}  for 91 days, between Monday, 29 July 2024, 00:00:00~UTC and Sunday, 27 October 2024, 23:59:59~UTC. Within the ROI, we collected a total of $6{,}881{,}633$ position-report messages from moving vessels (navigational status zero) and ship static data from $3325$ different Maritime Mobile Service Identity (MMSI) numbers associated with AIS-A vessels. Messages were stored with one-second timestamp precision, latitude and longitude with six decimal places, MMSI number, and the reported speed over ground of the position reports, as well as vessel type and destination information from the static reports.

\begin{figure}[ht!]
	\includegraphics[width=\textwidth, trim=0px 0px 0px 15px, clip]{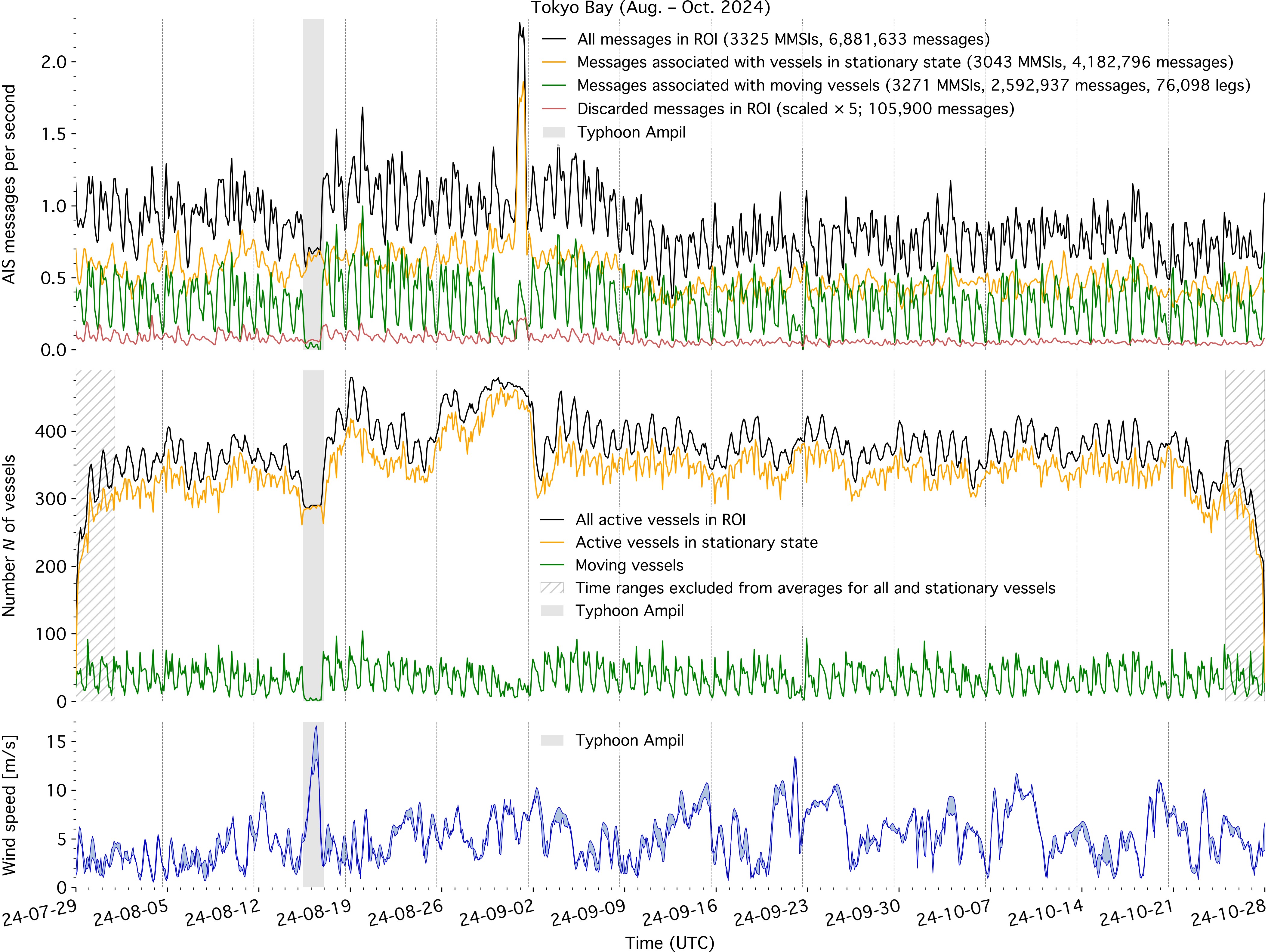}
	  \caption{
Timeline 
 of AIS-A message counts (top panel), inferred number of vessels operating in the ROI at each time (middle panel), and wind conditions in the bay (bottom panel) during the analysis period. The vertical dotted lines indicate the beginning of each Monday in local time (UTC+9). The black curves in the top and middle panels represent the sum of the other curves. The hatched gray-shaded intervals in the middle panel are excluded from the time-averaged estimates of the total number of vessels and of the stationary vessels in the ROI. Wind conditions are shown as the range of the initial values provided by the \citet{Copernicus2024} and \citet{GFS2024} forecast models, averaged over the ROI area. The passage of Typhoon Ampil along the Japanese coast on 16 August 2024 is indicated by the gray-shaded area in all three panels.} 
    \label{fig:ais_signals_time_tokyo}
\end{figure}

Figure \ref{fig:roi_data_map_tokyo_both} shows the message distribution in space, and Figure \ref{fig:ais_signals_time_tokyo} (top panel, black curve) shows the distribution in time. 

Figure~\ref{fig:roi_data_map_tokyo_both}a also displays four AIS receiver stations. Three of them---the Urayasu radio tower, the Yokohama Port Symbol Tower, and the Cape Kannon headquarters of the Tokyo Maritime Traffic Information Service (red radio tower symbols from top to bottom)---are reported by the network. However, we assume that the majority of messages are received through a station not listed by the network (``Tokyo receiver''), as indicated by the results in \Cref{sec:results_basestation}. Its published position is displayed by a red disk, and its position inferred from shadowing patterns is marked by a green diamond with a surrounding 68\% containment ellipse (see Section~\ref{sec:results_basestation} for details). Figure~\ref{fig:roi_data_map_tokyo_both}a additionally shows that the average positions where the first (black circle) or last (black diagonal cross) AIS messages from inbound or outbound vessels are received lie outside the hatched transit area. Properly accounting for fading receiver coverage toward the outer bay is one of the main analysis challenges addressed in \Cref{sec:data_processing}.\vspace{-3pt}

\begin{figure}[ht!] 
    \includegraphics[width=0.5\textwidth, trim=-4.5px 0px 0px 15px, clip]{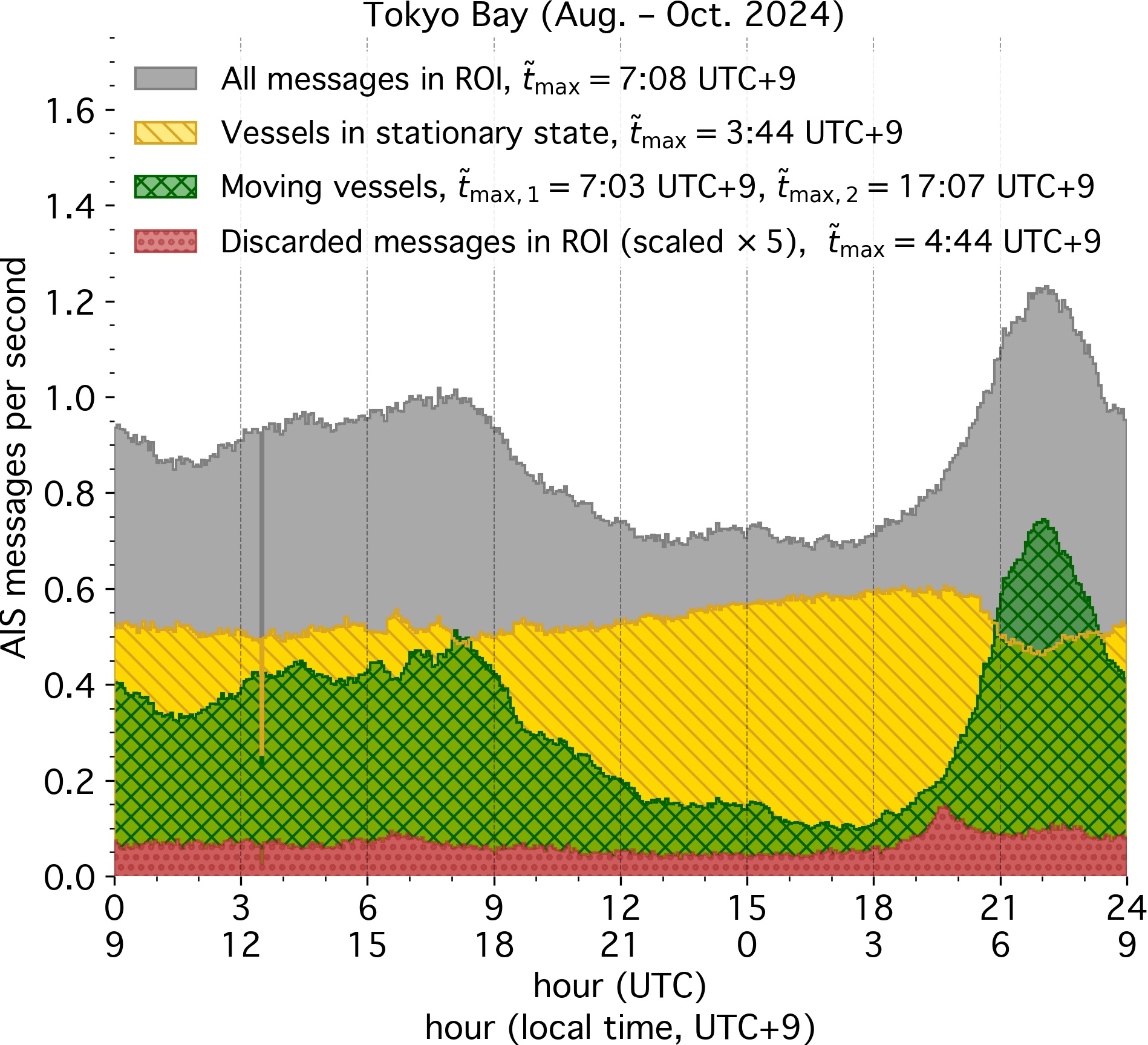}
    \caption{
    AIS message rate (same as Figure~\ref{fig:ais_signals_time_tokyo}, top) at different times of the day, as the average in 4-minute intervals over the 91-day analysis period. The times $\tilde{t}_\text{max}$ are the modes of maximum activity. The gap at 3:30 UTC is a data provider feature, and the peak in discarded messages at 4:44 UTC+9 is caused by isolated movements within the transit area (hatched area in Figure~\ref{fig:roi_data_map_tokyo_both}a).
    }
    \label{fig:hist_times_messages_cart_tokyo}

\vspace{17pt}

    \includegraphics[width=.5\textwidth, trim=0 0 0 15px, clip]{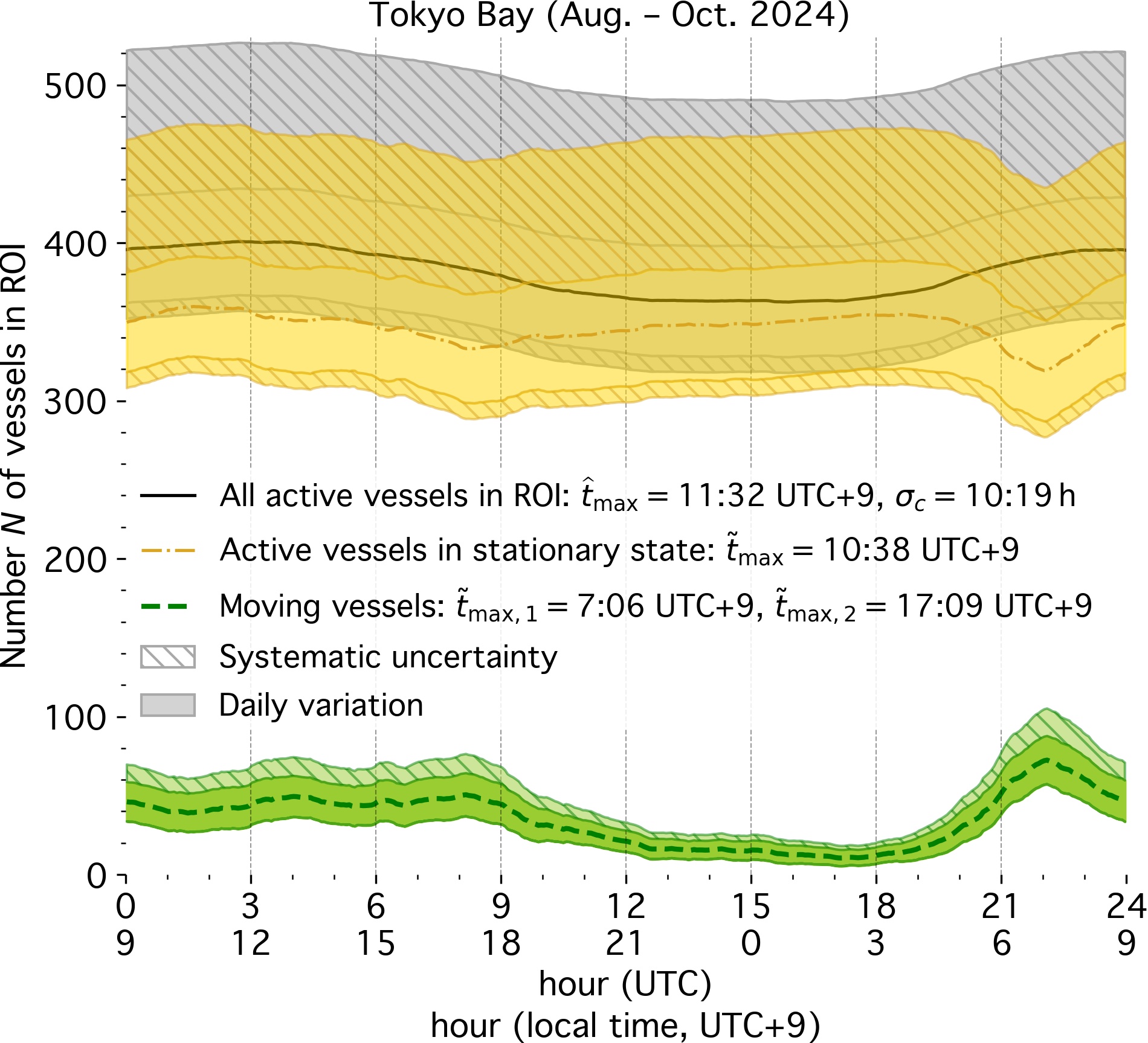}
    \caption{Vessel activity (same as Figure~\ref{fig:ais_signals_time_tokyo}, middle) in the bay depending on the daytime. The times $\hat{t}_\text{max}$ and $\sigma_\text{c}$ denote the circular mean and standard deviation \citep{Mardia1999}. Untracked and AIS-B vessels dominate systematic uncertainties (hatched bands; see discussion in Section~\ref{sec:discussion}).}
    \label{fig:hist_times_vessels_cart_tokyo}
\end{figure}

The message count over time (black curve in the top panel of Figure~\ref{fig:ais_signals_time_tokyo}) shows substantial fluctuations throughout each day and a slight decline after September, which is not reflected in the count of reconstructed vessels (middle panel). Also notable is a sharp, 12 h spike in signal activity on 1 September 2024, which is likewise absent from the reconstructed vessel count. The passage of Typhoon Ampil near Tokyo Bay on August 16 caused nearly all vessel motion in the bay to cease (gray-shaded areas). Figure~\ref{fig:hist_times_messages_cart_tokyo} (raw data) and Figure \ref{fig:hist_times_vessels_cart_tokyo} (results shown earlier in the paper, for comparison with the raw data) present the same data as in Figure~\ref{fig:ais_signals_time_tokyo} folded onto a daily cycle.

\subsection{AIS Data Processing and Vessel Status Classification}
\label{sec:data_processing}

We processed the AIS data using the same algorithm presented in our previous work \cite{Huetten2025a} to clean the data of erroneous messages, classifying the messages as belonging to either moving or stationary vessel phases, and creating simplified representations of the vessel movements over time (details are provided in Appendix~\ref{app:data_processing}). Figure~\ref{fig:ais_signals_time_tokyo} (top) shows the temporal distribution of messages separated and classified as from moving vessels (green), stationary vessels (yellow), and discarded messages (red). The classification assigned nearly all messages during the typhoon on 16 August 2024 to stationary vessels. Also, it associated the spike on 1 September 2024 with stationary vessels, cleaning this artifact from the movement periods (the spike affected both position and static-report messages).

\begin{figure}[ht!]

\includegraphics[width=0.5\textwidth]{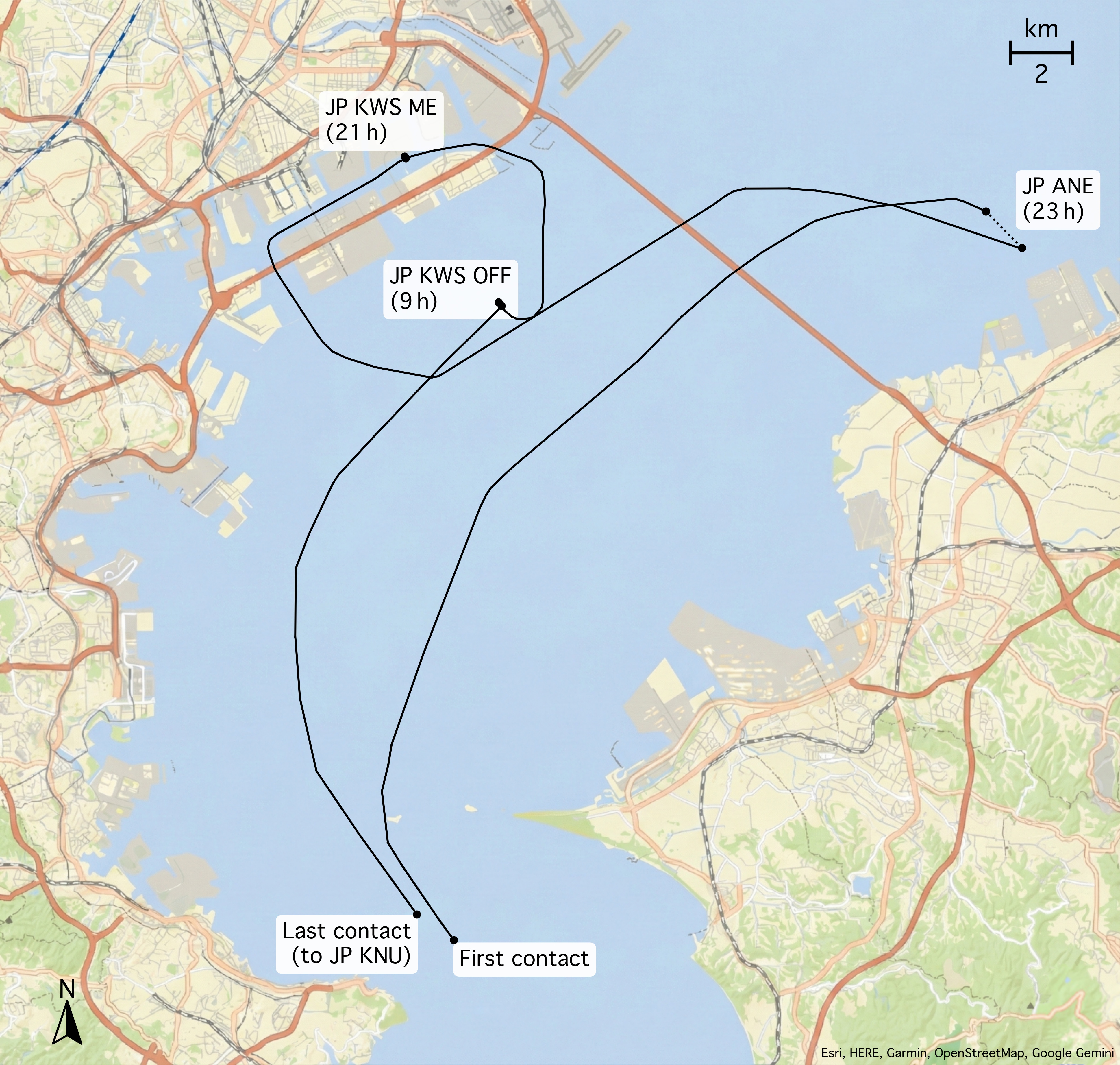}
    \caption{
    Recorded 
 journey of a liquefied petroleum gas (LPG) tanker (\SI{}{47{,}000} GT, \SI{230}{\meter} length). After the first AIS contact, the reported destination is Anegasaki port (JP$\,$ANE). Near Anegasaki, the signal is lost for \SI{23}{\hour}, after which the vessel proceeds to Kawasaki ME berth (JP$\,$KWS ME), anchors outside the harbor (JP$\,$KWS OFF), and finally departs toward Kinuura port (JP$\,$KNU) outside Tokyo Bay.
    } 
    \label{fig:lost_signal_example}
\end{figure}

A major challenge in analyzing our Tokyo Bay dataset was interpreting temporary gaps in a vessel's AIS message stream---that is, determining whether a vessel left and re-entered the bay during the gap or remained moored within the bay. As shown in Figure~\ref{fig:roi_data_map_tokyo_both}a, receiver coverage gradually diminishes toward the bay's opening, and first or last contact with entering or exiting vessels often occurs deep inside the bay. Even if a vessel remains inside the bay, its AIS data stream may be interrupted for two reasons: First, signal loss can occur within the bay due to the inhomogeneous receiver coverage. Second, vessels may switch off their AIS transceivers while moored, despite being required to keep them active \citep{Emmens2021}. Figure \ref{fig:lost_signal_example} illustrates this using a vessel journey from our dataset that consisted of four legs: The first signal was received at the transition to the inner bay, beyond our defined transit area. The signal was then lost again within the bay, likely due to a receiver coverage gap, before the vessel reached its first destination, Anegasaki, indicated by the vessel using the UN/LOCODE port code JP$\,$ANE \cite{JCG2023}. After 23 h, the signal resumed as the vessel proceeded to its second destination, JP$\,$KWS ME (Kawasaki ME berth). From 20 min after arrival, no signal was received for 21 h, until \SI{3}{\minute} before departure on the third leg. During this period, we confirmed that no position reports indicating moored status were received, suggesting that the AIS transceiver was likely switched off while mooring. At the third stop at JP$\,$KWS OFF, static reports were received at irregular intervals, on average every \SI{40}{\minute}. Although not used for the analysis, we confirmed that during this nine-hour mooring period, position-report messages with status 1 (``at anchor'') were also irregularly received by the network approximately every 10 min. On the final leg, the signal was lost shortly before the vessel exited inner Tokyo Bay.

\begin{figure}[ht!]

\includegraphics[width=0.5\textwidth]{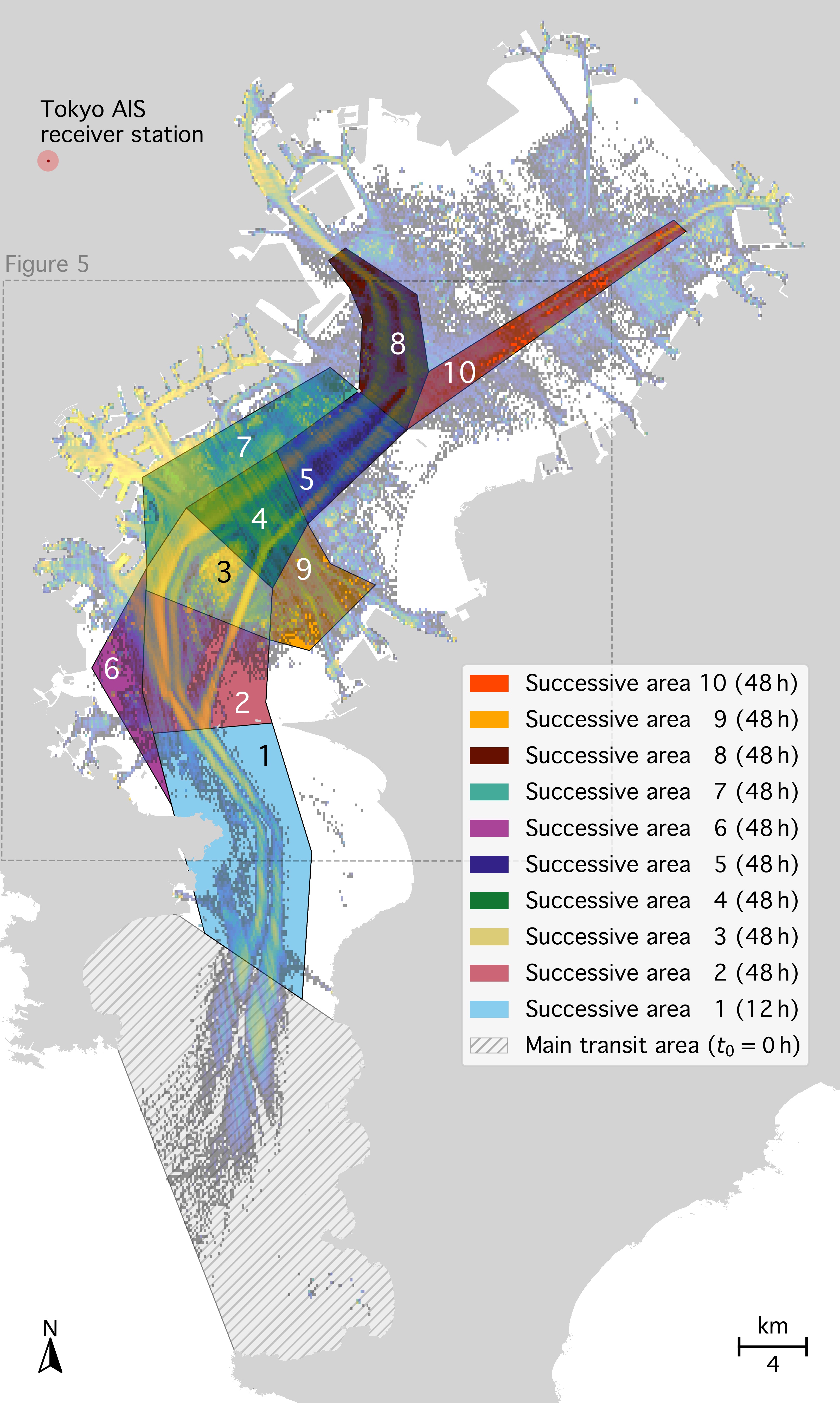}
    \caption{
    Main and successive transit areas within the Tokyo Bay ROI. We use successive areas 1 to 8 by default (\textit{df}).     To assess systematic uncertainties, we define a \textit{hi} (high) case using only successive areas 1 to 5. We add areas 9 and 10 to estimate the \textit{low} case. For validation purposes, we also cut through Nakanose anchorage under label `3' (see Section~\ref{sec:discussion}). We assume that receiver coverage decreases with distance from the Tokyo receiver (red disk). The message data from Figure~\ref{fig:roi_data_map_tokyo_both}a are shown in the background. The extent of Figure~\ref{fig:lost_signal_example} is indicated by the dashed frame.
    } 
    \label{fig:roi_exit_areas_tokyo}
\end{figure}

We used the transit-area approach introduced in \cite{Huetten2025a} to distinguish mooring and temporary absences from the ROI in cases of poor data coverage. To account for the peculiarity of this dataset, in particular the gradually diminishing receiver coverage, we introduced successive ``constrained'' transit areas following the shipping routes within Tokyo Bay, in addition to the main transit area defined in Figure~\ref{fig:roi_data_map_tokyo_both}a, with a progressively looser constraint $t_0$ the farther a vessel was inside the bay: if a leg ended or started in any of these areas, we treated the vessel as temporarily absent from the ROI only if (i) its period of stay was covered by fewer than four received messages, and (ii) the stay was longer than 
\begin{align}
    t_\text{thr} = t_0\times \left(\mathrm{max}(v_\text{exit}, v_\text{entry})\,/\,10\,\mathrm{kn}\right)^{-4}\,,
    \label{eq:t_thresh}
\end{align}    
where $v_\text{exit}$ is the last observed speed of the previous leg, $v_\text{entry}$ is the first speed value of the following leg, and $t_0=0,\,12,\,48\,\SI{}{\hour}$ is defined separately for each area, as shown in Figure \ref{fig:roi_exit_areas_tokyo}. Equation~(\ref{eq:t_thresh}) is the same as that introduced in \cite{Huetten2025a}. By default, we added the successive areas 1 to 8 to the vessel status classification; further area subdivision of areas 2--10, all with $t_0=\SI{48}{\hour}$, was used to assess uncertainties in the classification, as discussed in Section~\ref{sec:discussion}. 

The values of $t_0$ were motivated as follows: Outside the fairways, vessels are observed to generally travel at speeds $\lesssim\,$\SI{10}{kn}. In the outer bay, vessels remain below the maximum allowed speed of \SI{12}{kn} \citep{JCG2023}, while speeds on the inner bay lanes are higher \citep{Shimizu2020}. In addition, high-speed vessels operate at speeds $\gtrsim \SI{30}{kn}$ throughout the bay. For such observed or mandated speeds, Equation~(\ref{eq:t_thresh}) gives $t_\text{thr}^{\SI{10}{kn}}=t_0$, $t_\text{thr}^{\SI{12}{kn}}\approx 0.5\,t_0$, $t_\text{thr}^{\SI{15}{kn}}\approx 0.2\,t_0$, and $t_\text{thr}^{\SI{30}{kn}}\approx 0.01\,t_0$. So, high-speed ferries between Tokyo and Oshima Island, which may leave and return to the bay within two hours, are identified as absent by Equation~(\ref{eq:t_thresh}) for message gaps $\gtrsim\,$\SI{9}{\minute} (area 1) or $\gtrsim\,$\SI{36}{\minute} (areas 2--10). In contrast, vessels anchoring in the inner bay are only falsely classified as absent if their transceivers remain switched off for longer than \SI{48}{\hour}. The uncertainty discussion in Section~\ref{sec:discussion} quantifies the sensitivity of the derived vessel metrics to changing the here-defined transit-area parameters. In particular, it confirms that a threshold of \SI{48}{\hour} correctly identifies most anchoring vessels, i.e., a negligible number of vessels anchor in the bay with their AIS transceivers switched off for longer than \SI{48}{\hour}.

\subsection{Computation of Metrics and Berth Identification}
\label{sec:vessel_metrics}

Based on the reconstructed vessel movements and classified absence or mooring periods, we computed the vessel count within the bay and transit rates into or from the bay. Also, we derived gridded vessel density, speed, and course direction maps at a resolution of $(\Delta\varphi,\,\Delta\lambda)=(1'',\,1'')$, corresponding to a cell height  of $h=\SI{30.82}{\meter}$ in the latitude direction and widths $w$ between \SI{25.4}{\meter} ($\varphi=34.9^\circ$) and \SI{25.1}{\meter} ($\varphi=35.7^\circ$) in the longitude direction. The vessel density map was also used to identify berth and anchorage locations. The adopted density-thresholding approach constitutes an alternative to previous, mostly cluster-based methods for this purpose \citep{Yan2022,AsianDevelopmentBank2023,Zhang2023b,Iphar2024,Qiang2025,Hadjipieris2025}.  For all metric computations, we adopted the same methodology introduced in our previous work \cite{Huetten2025a}, with the parameters adjusted for the analysis given in Appendix~\ref{app:data_processing}.


\subsection{Receiver Localization}
\label{sec:methods_basestations}

The urban topography of the Tokyo Metropolitan Area causes sharply varying signal occlusions due to tall buildings between the transmitting vessels and receivers. Such radio shadowing is usually observed in natural topography \citep{Shelmerdine2015,Mazzarella2017,Jung2024} but can also occur at smaller structures \citep{Last2015}.  This allows us to infer the precise location of some of the network's receiver stations in the area from the vessel AIS data, as the shadow patterns trace back to the receiver stations at focal points of the extended shadow lines. In Tokyo Bay, the data stream appears to be dominated by a single receiver station about \SI{10}{\km} northwest of Tokyo Bay (``Tokyo receiver''), creating a pronounced spatial structure in received vessel positions, as shown in Figure~\ref{fig:roi_data_map_tokyo_both}. A second, much weaker pattern was also found in the data, hinting at a receiver station west of Yokohama (``Yokohama receiver''). 
 
\begin{figure}[ht!]
   
    \begin{minipage}{0.49\textwidth} %
    \vspace{-3.31cm}
    \subfloat[
   \label{fig:roi_pattern_tokyo}
    ]{\includegraphics[width=\textwidth]{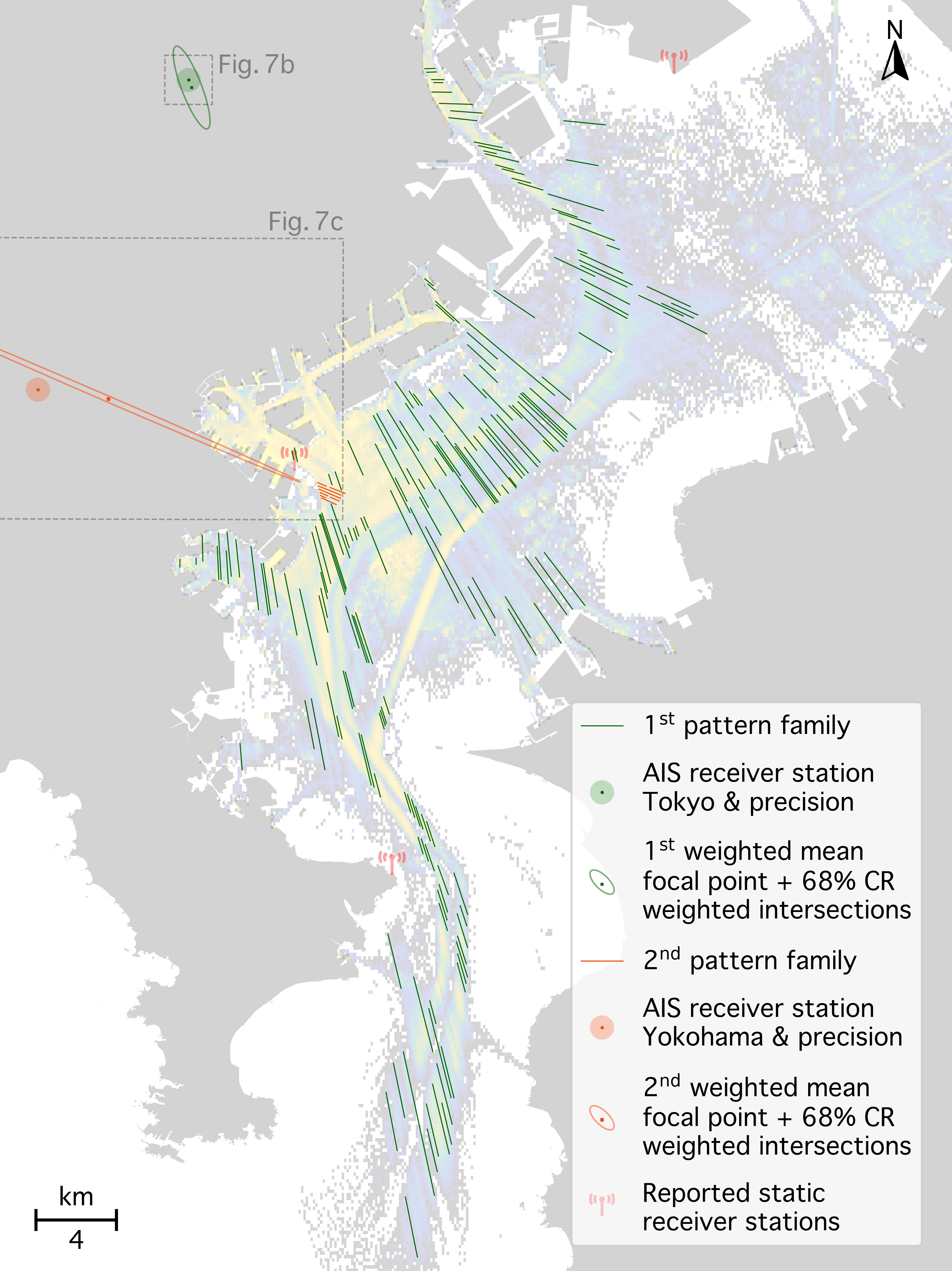}}
  \end{minipage}%
\hfill
    \begin{minipage}{0.49\textwidth}
           
    \subfloat[
\label{fig:mysterious_receiver_close}
]{\includegraphics[width=\textwidth]{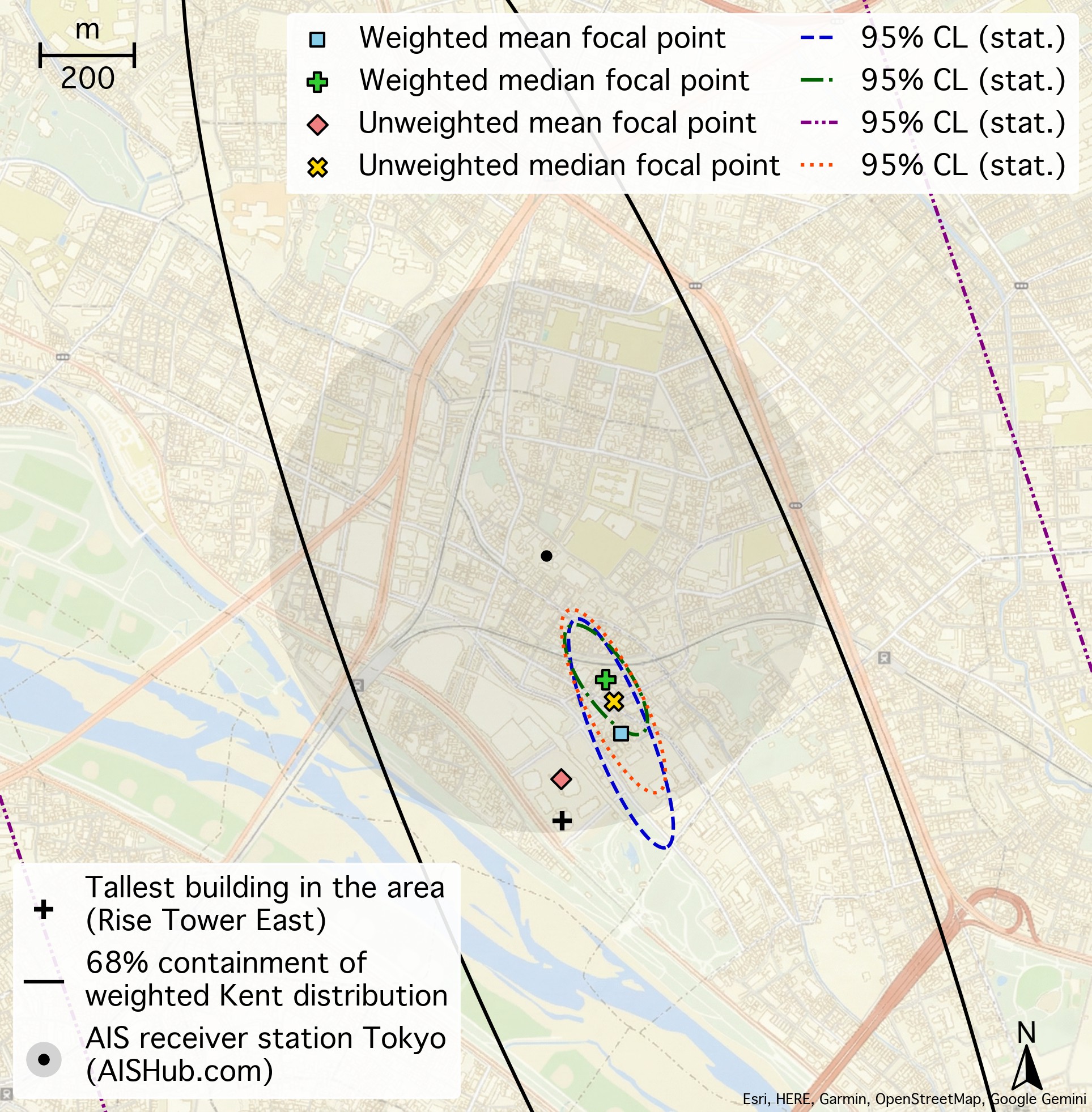}}
\vspace{-0.2cm}
    \subfloat[
    \label{fig:known_receiver_close}
 ]{\includegraphics[width=\textwidth]{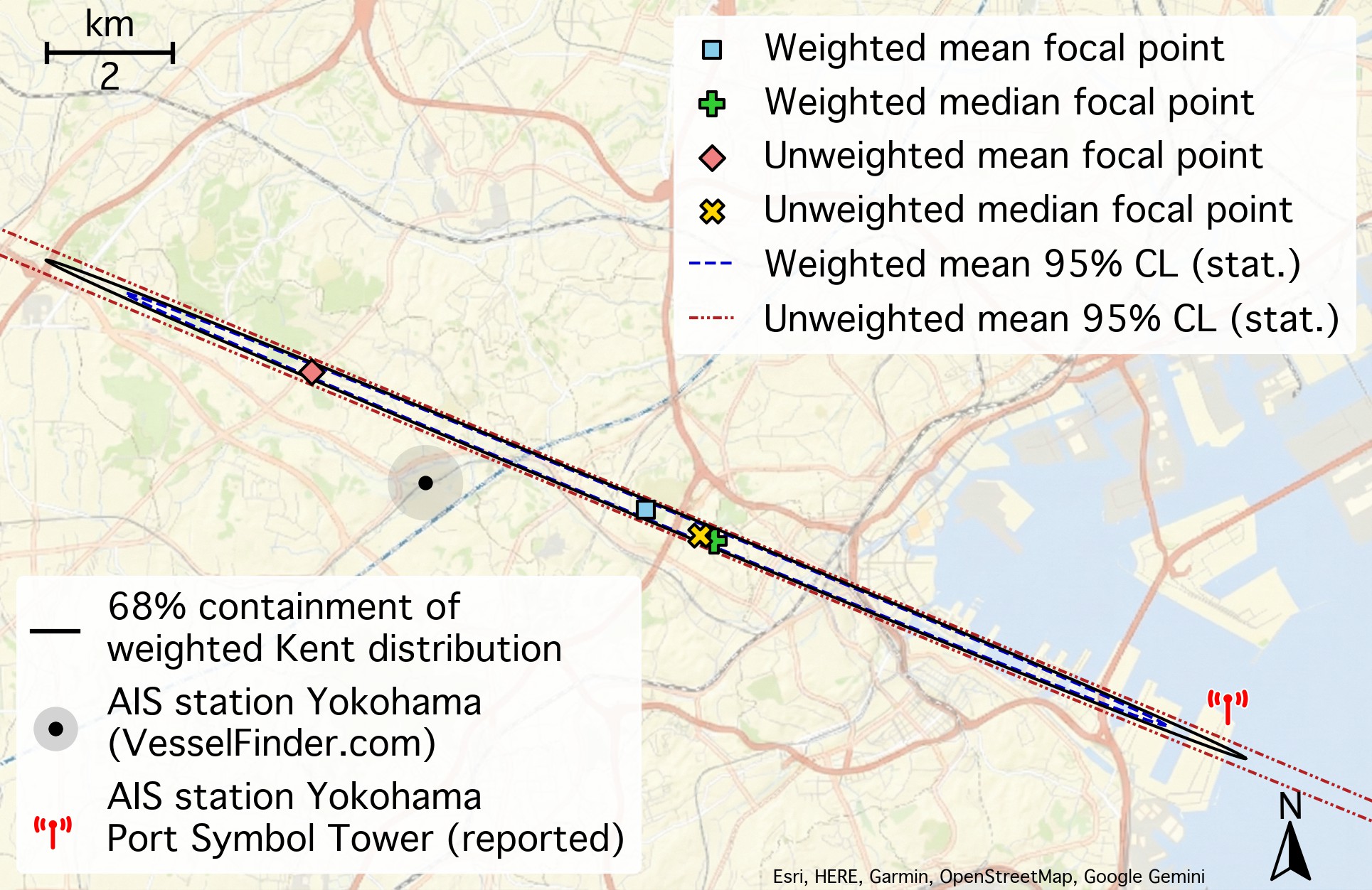}}
     \end{minipage}
    \caption{
AIS 
 receiver stations inferred from AIS message shadows in Tokyo Bay.
 (\textbf{a})  A total of 198  geodesic segments were fitted to the radio shadows in the AIS message distribution in Tokyo Bay, with 190 segments associated with the Tokyo receiver  (green lines; close-up in the upper-right panel), and 8 associated with the Yokohama receiver (orange lines; close-up in the lower-left panel). Neither of the inferred receiver positions (ellipses and circles) aligns with the stations reported by the network (red-shaded tower symbols in the left panel).
 (\textbf{b}) Position estimates of the Tokyo receiver station. The black dot shows AISHub~\cite{AISHub2025} station 3752.  (\textbf{c}) Position estimates of the Yokohama receiver station. The black dot indicates VesselFinder~\cite{VesselFinder2025} station 5552 (JR1CAD).
  Both panels on the right show the mean (squares and diamonds) and median (upright and diagonal crosses) positions of the focal points of weighted (blue and green) and unweighted (red and yellow) sets of  radio shadows, along with their confidence ellipses (no median position confidences are computed for the Yokohama receiver due to low statistics). The solid-line ellipses show the 68\% containment range (CR) of all segment intersections. The dots and disks mark the receiver positions from \cite{VesselFinder2025,AISHub2025} within an uncertainty radius of $\Delta\varphi=0.005^\circ$ (\SI{550}{\meter}). Numeric values are listed in Table~\ref{tab:receiver_positions}.
    } 
    \label{fig:inferred_receivers_close}
\end{figure}

To find the focal points, we fit 198 geodesic sections onto the AIS message density, binned into \SI{130}{\square\meter} pixels, as shown in Figure \ref{fig:inferred_receivers_close}a. We associated 190 of these sections with the Tokyo receiver (green lines in Figure~\ref{fig:inferred_receivers_close}a) and eight with the Yokohama receiver station (orange lines). We calculated all 17955 (28) pairwise geodesic intersections for the Tokyo receiver (Yokohama receiver) and used the 
{\tt sphstat} software package \citep{Hachabiboglu2023} to remove 48 (1) intersection outliers, following \citet{Fisher1981}, at the 95\% confidence level ($\alpha=0.05$), assuming a symmetric von~Mises–Fisher distribution \citep{Fisher1953}. For both sets, we assigned a weight to each intersection based on the sine of the enclosing angle $\gamma$. For the Tokyo receiver, we set to zero the weights of 937 intersections with weight values below 0.02 ($\gamma<1.1^\circ$). For the Yokohama receiver, the remaining 27 intersections were all nearly parallel ($0.2^\circ\leq \gamma \leq 3.4^\circ$), and all intersections were considered scaled by their weight.  
 
 We calculated the spherical mean \citep{Fisher1993a} and median \citep{Fisher1985} focal points for both receiver stations with {\tt sphstat} and estimated the parameters of a Kent distribution \citep{Kent1982}, the analogue of a bivariate Gaussian distribution on the sphere. The semi-axes $a = \sigma_2/\mu$ and $b = \sigma_3/\mu$ define an ellipse-like region 
\begin{equation}
    \frac{\nu^{*2}_2}{a^2} + \frac{\nu^{*2}_3}{b^2} = -2\log(1-p)
\end{equation}
around the spherical mean $\boldsymbol{\nu}^*\in\Omega_3$, which encloses the distribution's $p\times100\%$ containment range. The parameters $\mu$, $\sigma_j$, and $\nu_j^*$ were defined according to \citet{Kent1982} (Equations~(3.8) and (6.2)). In the limit  $\lim_{(a,b)\rightarrow 0} (a,b) =(\sigma_x,\sigma_y)$, one obtains the standard errors of a bivariate Gaussian distribution. We tested that inferring the weighted mean focal point on a tangential plane, assuming a bivariate Gaussian distribution, results in only a \SI{20}{\cm} difference for the mean position of the Tokyo receiver, but in 3\% and 15\% relative differences for the semi-axis lengths $a$ and $b$, respectively.

We estimated the confidence ellipse for the spherical mean using a Hotelling t-squared statistic \citep{Hardle2019}, 
\begin{equation}
    \frac{\nu^{*2}_2}{a^2} + \frac{\nu^{*2}_3}{b^2}\; \leq\; \frac{m\,F_{m,\,n-m,\,p}}{n-m} \; \overset{\substack{n\rightarrow\infty \\ m=2}}{\lesssim} \;\frac{-2\log(1-p)}{n-2}\,,
    \label{eq:confidence_ellipse}
\end{equation}
where $F_{m,\,n-m,\,p}$ is the inverse cumulative distribution function of the {\it F}-distribution \citep{Draper1998} at the $p=1-\alpha$ percentile, $m=2$ is the number of spatial dimensions, and $n$ represents the number of independent measurements, taken as the number of fitted segments. For the Tokyo receiver, we used $F_{2,\,188,\,0.95}=3.044$, and for the Yokohama receiver, we used $F_{2,\,6,\,0.95}=5.143$. Similarly, we estimated the confidence ellipse of the spherical median \citep{Fisher1985}.

\section{Results}
\label{sec:results}

\subsection{Vessel Activity in Tokyo Bay}
\label{sec:results_vessels}

\subsubsection{Average Vessel Count and Traffic 
}
{
Between August and October 2024, we found, on average, 381 vessels simultaneously operating in Tokyo Bay, of which more than 90\% were engaged in stationary activities such as berthing, fishing, or anchoring. On average, 35 vessels were actively moving between berths, ports, or anchorages, or in the process of entering or leaving the bay. During the observed period, we found an average of 293 entries or exits per day.

Table \ref{tab:vessel_numbers} categorizes vessels in Tokyo Bay into six types according to self-reported activity (using the code mapping in Table~A2  in \cite{Huetten2025a}). Commercial cargo shipping dominates (41\%), followed by liquid bulk or tanker vessels (24\%), and 23\%
 of the activity is comprised of pilot, tug, or construction vessels. Fishing boats and passenger transportation together account for less than 10\% of total activity in the bay.

\begin{table}[ht!]
\caption{Average number $N$ of vessels active in Tokyo Bay (August
--October 2024). The first error terms denote the statistical fluctuations between different moments of observation. The second error terms describe the systematic error due to uncertainty about whether vessels leave and re-enter or remain present unobserved (not reflected in the moving vessel count), as well as additional untracked or AIS-B vessels (upper bounds only). Gross-tonnage (GT) data could be retrieved for $99\%$ of vessels in the analysis with an International Maritime Organization (IMO) number.}
\label{tab:vessel_numbers}
\scriptsize
\begin{adjustwidth}{-1.5cm}{0cm}
\newcolumntype{C}{>{\centering\arraybackslash}X}
\begin{tabularx}{\fulllength}{Ccccccc}
\toprule
\textbf{Vessel Type} & \textbf{All Active Vessels} & \textbf{Moving Vessels} & \textbf{Stationary Vessels}
 & \textbf{Percentage (All)} & \textbf{Percentage (Moving)} & \textbf{Percentage (Stationary)} \\
\midrule

All vessels in ROI &

$381 \pm 14\,{}^{+84}_{-10}$ &
$ 35 \pm 17\,{}^{+\;\,8}_{-\;\,0}$ &
$346 \pm \;\,9\,{}^{+84}_{-10}$ &
$100\%$ &
$\;\,9\%$ &
$91\%$ \\\midrule

Passenger, high-speed &

$\;\;\,12 \pm \;\,1\,{}^{+2.0}_{-0.3}$ &
$ \;\;2 \pm \;\,1\,{}^{+0.4}_{-0.0}$ &
$\;\;10 \pm \;\,1\,{}^{+2.0}_{-0.3}$ &
$\;\,\;\,3\%$ &
$\;\,6\%$ &
$\;\,3\%$ \vspace{3pt}\\

Law enforcement, military &

$\;\;5.0 \pm 0.1^{+1.0}_{-0.1}$ &
$\,0.1 \pm 0.1\,{}^{+0.0}_{-0.0}$ &
$\;\,4.9  \pm 0.0^{+1.0}_{-0.1}$ &
$1.3\%$ &
$0.3\%$ &
$1.4\%$ \vspace{3pt}\\

Cargo &

$   158 \pm \;\,9\,{}^{+30}_{-\;\,4}$ &
$    15 \pm \;\,7\,{}^{+\;\,3}_{-\;\,0}$ &
$   144 \pm \;\,7\,{}^{\,+30}_{-\;\,4}$ &
$41\%$ &
$42\%$ &
$42\%$ \vspace{1pt}\\

Pilot, tug, rescue, diving/dredging &

$\;\;\,87 \pm 0.4\,{}^{+25}_{-\;\,2}$ &
$10 \pm \;\,6\,{}^{+\;\,3}_{-\;\,0}$ &
$\;\,77 \pm \;\,6\,{}^{+25}_{-\;\,2}$ &
$23\%$ &
$28\%$ &
$22\%$ \vspace{1pt}\\

Tanker &

$\;\,93 \pm \;\,6\,{}^{+19}_{-\;\,2}$ &
$ \;\,8 \pm \;\,5\,{}^{+\;\,2}_{-\;\,0}$ &
$\;\, 85 \pm \;\,4\,{}^{+19}_{-\;\,2}$ &
$24\%$ &
$21\%$ &
$25\%$ \vspace{3pt}\\

Others, including fishing &

$\;\;\,26 \pm 0.3\,{}^{+15}_{-\;\,1}$ &
$\;\;\,\;1 \pm 0.5\,{}^{+0.5}_{-0.0}$ &
$\;\;\,25 \pm 0.2\,{}^{+15}_{-\;\,1}$ &
$\;\,7\%$ &
$\;\,3\%$ &
$\;\,7\%$ \\\midrule

All IMO vessels &

$283 \pm 12\,{}^{+62}_{-\;\,8}$ & 
$26 \pm 11\,{}^{+\;\,6}_{-\;\,0}$ &
$257 \pm \;\,7\,{}^{+62}_{-\;\,8}$ & 
$74\%$ &  
$73\%$ & 
$74\%$ \\[0.1cm]  

Vessels with GT $<10{,}000$ &

$231 \pm \;\,8\,{}^{+55}_{-\;\,6}$ &
$20 \pm \;\,9\,{}^{+\;\,5}_{-\;\,0}$ &
$211 \pm \;\,6\,{}^{+55}_{-\;\,6}$ &
$\,\geq$60\%$\quad$ & 
$\,\geq$56\%$\quad$ &
$\,\geq$61\%$\quad$ \\  

Vessels with GT $\geq10{,}000$ &

$\;\,52 \pm \;\,5\,{}^{+\;\,9}_{-\;\,1}$ &
$\;\,6 \pm \;\,3\,{}^{+\;\,1}_{-\;\,0}$ &
$\;\,46 \pm \;\,4\,{}^{+\;\,9}_{-\;\,1}$ &
$\,\geq$14\%$\quad$ &
$\,\geq$17\%$\quad$ &
$\,\geq$13\%$\quad$ \\

\bottomrule
\end{tabularx}
\end{adjustwidth}
\end{table}

 Figure \ref{fig:hist_times_vessels_cart_tokyo} (already shown in Section~\ref{sec:methods}) displays the vessel activity throughout the day. Vessel motion sharply peaks in the morning around 7 a.m. local time and is followed by a plateau throughout the day until around 17:30 (green dashed curve and bands). Traffic drops during the night and resumes around 4 a.m. the next morning. Notably, the number of stationary vessels peaks during the day at around 10:30 a.m. (yellow dashed-dotted curve and bands), with the overall count being highest around noon, nearly 10\% higher than during the night. This suggests that
  vessels tend to enter the bay during the day for (primarily commercial) activities and leave at night. 

We found few references to vessel counts within Tokyo Bay. In their worldwide analysis, \citet{Kroodsma2018}} identified an average of 2.6 fishing vessels in our ROI between 2012 and 2016, 
significantly fewer than our 26 vessels in that category. However, when restricting fishing activity to vessels reporting AIS code 30 (codes 1002 and 1003 were not present in our data), we counted 2.5 simultaneously active fishing vessels, closely matching the number from \cite{Kroodsma2018}.

\begin{figure}[ht!]
    \includegraphics[width=0.55\textwidth, trim=0px 0px 0px 15px, clip]{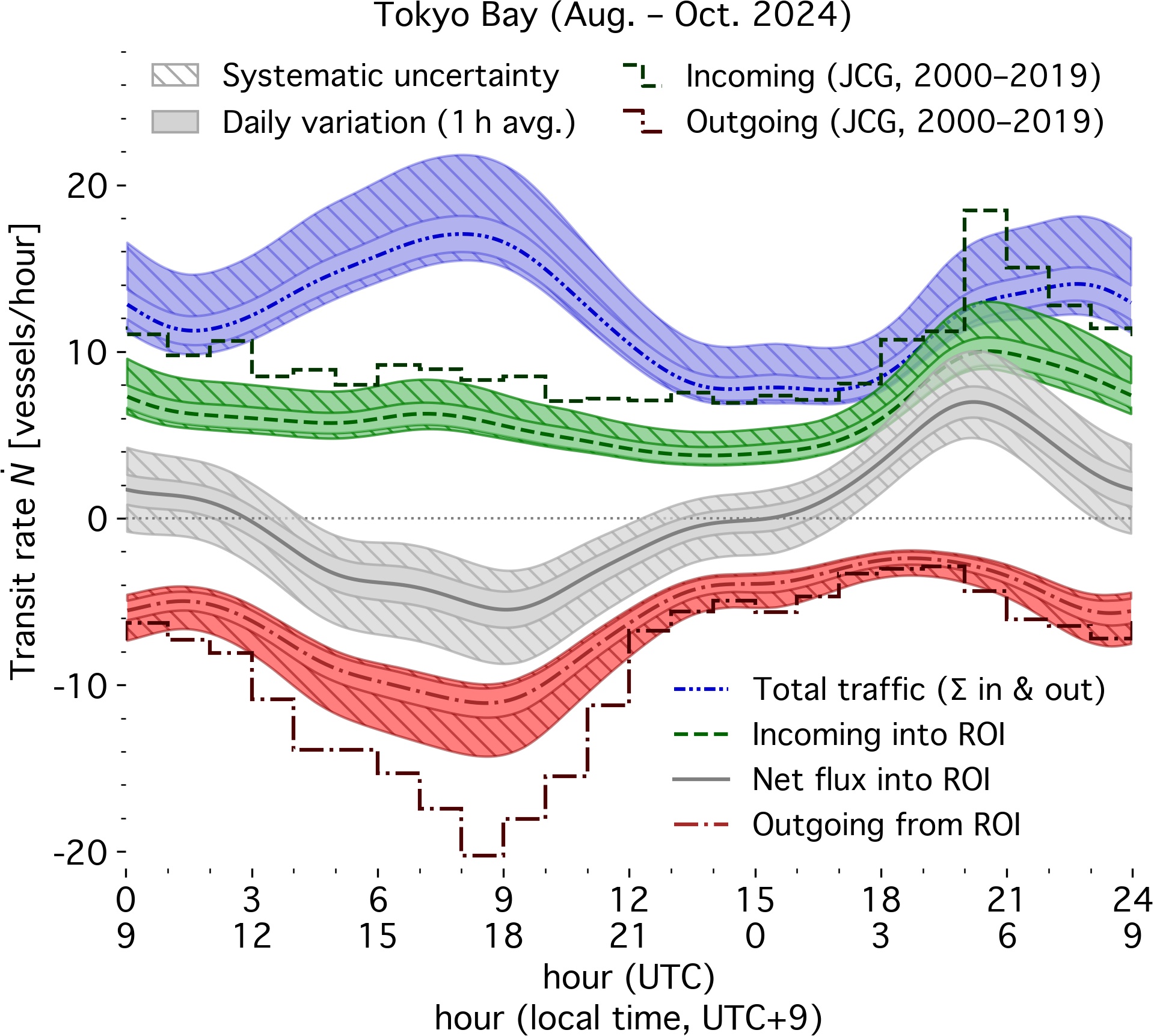}
    \caption{
    Hourly 
 transit rate of vessels entering or leaving Tokyo Bay (August--October 2024). Our results (smooth curves, solid daily-fluctuation bands, and hatched uncertainty bands) are compared with Uraga Channel traffic data from \cite{JCG2020} between 2000 and 2019 (green and red step curves).
    } 
    \label{fig:vessel_flux_tokyo}
\end{figure}

Figure \ref{fig:vessel_flux_tokyo} shows our results for transit traffic throughout the day (blue dashed-double-dotted curve and bands), further distinguishing inbound (green dashed curve and bands), outbound (red dashed-dotted curve and bands), and net incoming traffic (gray solid curve and bands). The net traffic $\dot{N}$ corresponds to the time derivative of the total vessel number $N$ (gray curve) in Figure~\ref{fig:hist_times_vessels_cart_tokyo}. The \citet{JCG2020} provides statistics on vessel passages through the Uraga Channel for 48 days between 2000 and 2019, on two or three consecutive days each year. The average from these days, corresponding to a total of 449 transits per day, is displayed as dashed green and dot-dashed red step curves. Our analysis reproduces this temporal pattern: inbound traffic peaks around 6~a.m. local time, outbound traffic shortly before 18:00, and overall transit traffic (blue curve) is highest between 16:00 and 17:00 in the afternoon. The same pattern was also observed in the 1990s \citep{Kobayashi1998}.

\begin{figure}[ht!]
    \includegraphics[width=0.6\textwidth]{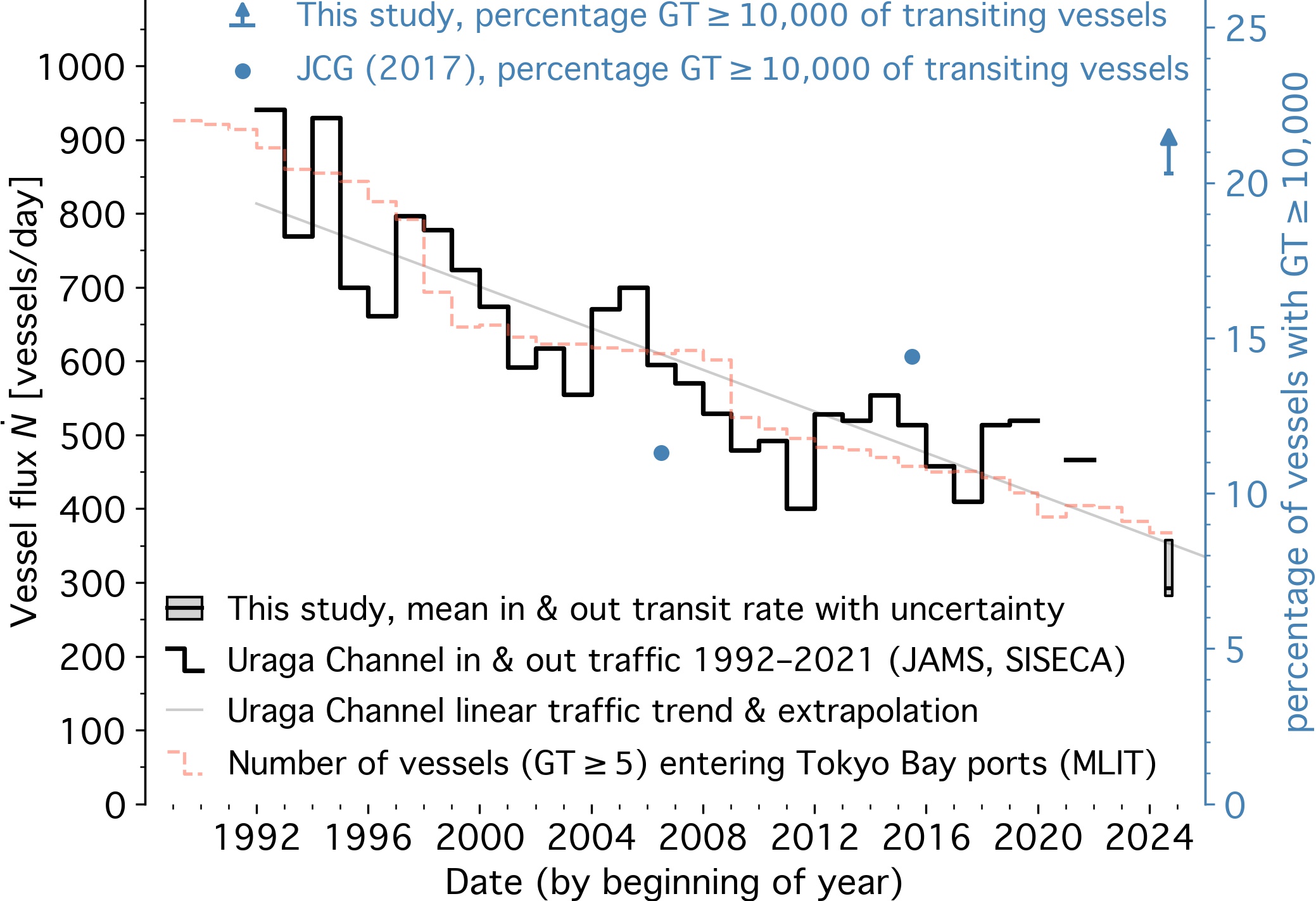}
    \caption{
Tokyo 
 Bay transit traffic over time (inbound and outbound; left black scale) and percentage of large vessels above 10,000 GT among this traffic (right blue scale). Past traffic is shown by data from \cite{JAMS2023,SISECA2018,SISECA2023} (black step curve) and its regression line (thin gray curve). Also, port calls in Tokyo Bay from \cite{MLIT2025} are shown for comparison (dashed orange step curve). Blue dots show the large-vessel percentages from \cite{JCG2017}. Our mean estimated traffic with systematic uncertainties (Section~\ref{sec:discussion}) is shown by the gray box, and our results on the lower limit of the large vessels by a blue arrow.
}
    \label{fig:vessels_historic_tokyo}
\end{figure}

The Japan Coast Guard also publishes yearly traffic averages  through the Uraga Channel zone, as indicated in Figure~\ref{fig:roi_data_map_tokyo_both}a (\citet{Kobayashi1998} precisely defines the ``line between Sea Forts Nr.$\,$2 and Nr.$\,$3''). Figure \ref{fig:vessels_historic_tokyo} illustrates the long-term trend between 1992 and 2021 based on \cite{JAMS2023,SISECA2018,SISECA2023} (black step curve; no data are available for 2020), along with a fitted linear trend (thin gray line). The data show a clear decline in traffic over the past two decades, and our results for 2024 align with this trend within systematic uncertainties. For reference, Figure~\ref{fig:vessels_historic_tokyo} also includes data from the Japanese Ministry of Land, Infrastructure, Transport, and Tourism \cite{MLIT2025} on the average daily port calls per year of vessels larger than 5~GT in the ports of Tokyo Bay (Tokyo, Kawasaki, Yokohama, Yokosuka, Chiba, Kisarazu; orange step curve; Funabashi and Katsunan ports are not listed by \cite{MLIT2025}). However, port call counts are not directly comparable to transit traffic, as external vessels may make multiple port calls during a single Tokyo Bay visit, and some vessels operate exclusively within the bay. 

The \citet{JCG2017} reported that the decline in vessel numbers was accompanied by an increase in vessel size. The proportion of transiting vessels exceeding a gross tonnage (GT) of 10,000~GT rose from 11\% in 2006 to 14\% in 2015 (blue dots in Figure~\ref{fig:vessels_historic_tokyo}). Our data from three months in 2024 confirm this trend and suggest further acceleration, with at least $20\%$ of transiting vessels above 10,000 GT (Table~\ref{tab:vessel_numbers_flux} and blue arrow in Figure~\ref{fig:vessels_historic_tokyo}). The corresponding mean GT per vessel was $11{,}864^{+275}_{- \;\,54}$, and an extrapolated yearly cumulative GT of at least $\left(1.085^{+0.013}_{- 0.003}\right)\times10^9$ moved into or out of Tokyo Bay in 2024.

Table \ref{tab:vessel_numbers_flux} further categorizes inbound and outbound vessels. The number of auxiliary pilot and construction vessels was lower than their average share within the bay, and nearly 90\% of transit traffic consisted of commercial vessels.

\vspace{-6pt}

\begin{table}[ht!]
\caption{Number 
 $\dot{N}$ of vessels entering or leaving Tokyo Bay each day (transit rate) between August and October 2024. 
The first errors show the daily fluctuations, and the second the analysis uncertainty.
}
\label{tab:vessel_numbers_flux}
\begin{tabular*}{\textwidth}{lcc}
\toprule
\textbf{Vessel Type} & \textbf{Transits/Day,  $\boldmath{\dot{N}}$} & \textbf{Percentage}  \\\midrule
 All Vessels & 
  $293\pm22\,{}^{+65}_{-10}$ & $100\%\;\;$  \\
\midrule
 Passenger, high-speed & 
 $\;\,16 \pm \;\;1\,{}^{+\;\,3}_{-\;\,1}$ & $\;\,5\%$\\[0.15cm]
 Law enforcement,  military & 
 $\;\;0.8 \pm 0.1\,{}^{+0.3}_{-0.0}$ & \hspace{-0.1cm}$<$1\%\\[0.15cm]
 Cargo & 
 $\,162\pm 12\,{}^{+\,34}_{-\;\,4}$ & $\;55\%$ \\[0.15cm]
 Pilot, tug, rescue, diving/dredging  & 
 $\;\,13 \pm \;\,1\,{}^{+\;\,6}_{-\;\,1}$ & $\;\,4\%$\\[0.15cm]
 Tanker & 
 $\;\,93\pm \;\,7\,{}^{+20}_{-\;\,5}$ & $32\%$\\[0.15cm]
 Others, including fishing& 
 $\;\;\;\,7 \pm \;\,1\,{}^{+4.7}_{-0.3}$ & $\;\,2\%$\\
 \midrule
   IMO vessels&
  $250\pm 18\,{}^{+61}_{-\;\,9}$  & $\;85\%$ 
 \\[0.2cm]
Vessels with GT $<10,000$&
  $191\pm14\,{}^{+51}_{-\;\,8}$   & \hspace{-0.1cm}$\geq$65\%$\;$
 \\[0.15cm]
Vessels with GT $\geq10,000$&
 $\;\,59 \pm \;\,4\,{}^{+12}_{-\;\,1}$  & \hspace{-0.1cm}$\geq$20\%$\;\;$\\
\bottomrule
\end{tabular*}
\end{table}

\subsubsection{Spatially Resolved Activity and Berths}
\begin{figure}[ht!]
\captionsetup[subfloat]{margin=2pt}

    \subfloat[    
  \label{fig:bearing_map_tokyo}]{
     \tikz[remember picture]{
      \node[inner sep=0pt,anchor=south west] (A) {
    \includegraphics[width=0.5\textwidth, trim=0px 0px 0px 15px, clip]{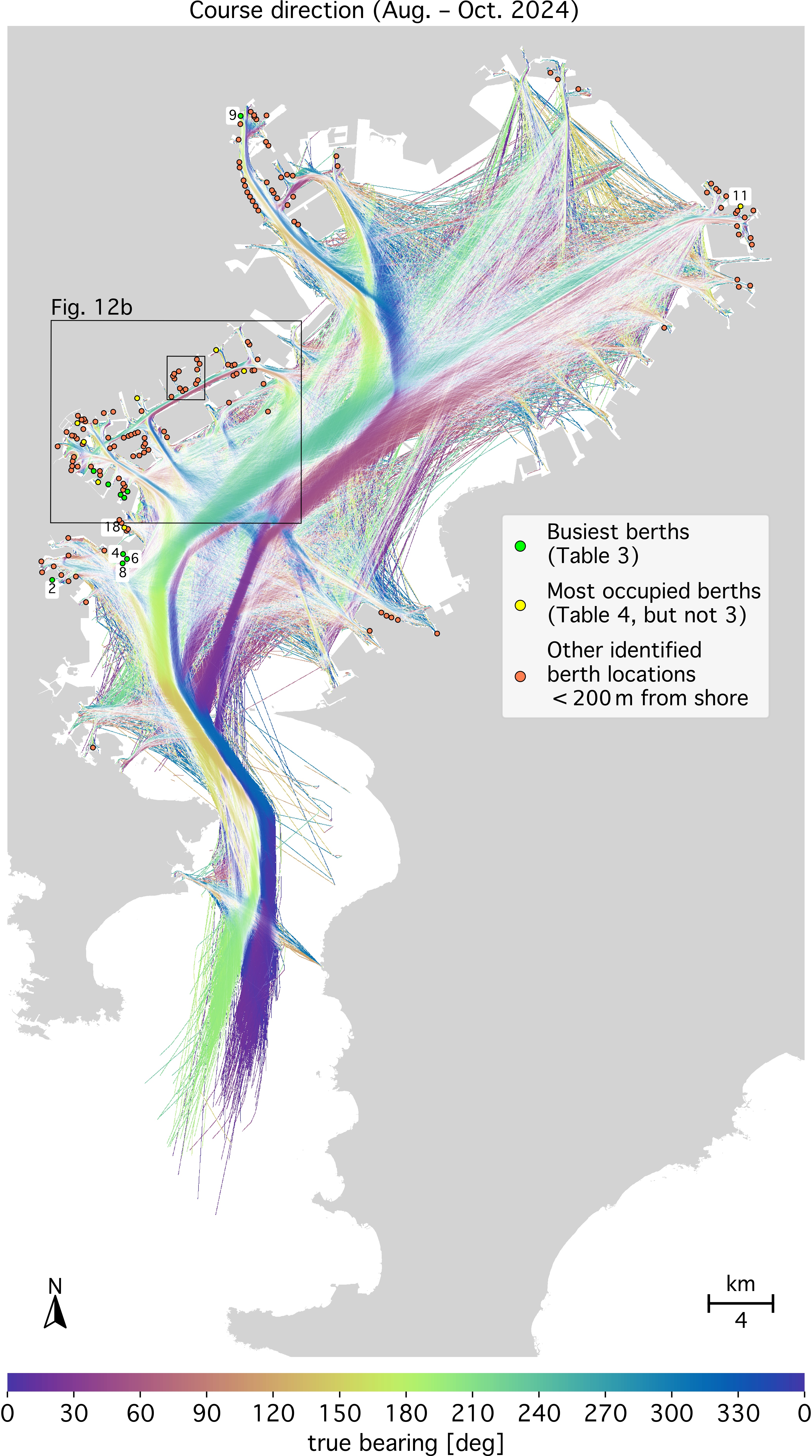}
    };
    }
}
\hfill
\raisebox{.599\height}{
 \subfloat[ 
    \label{fig:islands_closeup_tokyo}
 ]{
 \tikz[remember picture]{
  \node[inner sep=0pt,anchor=south west] (B){
    \includegraphics[width=0.47\textwidth]{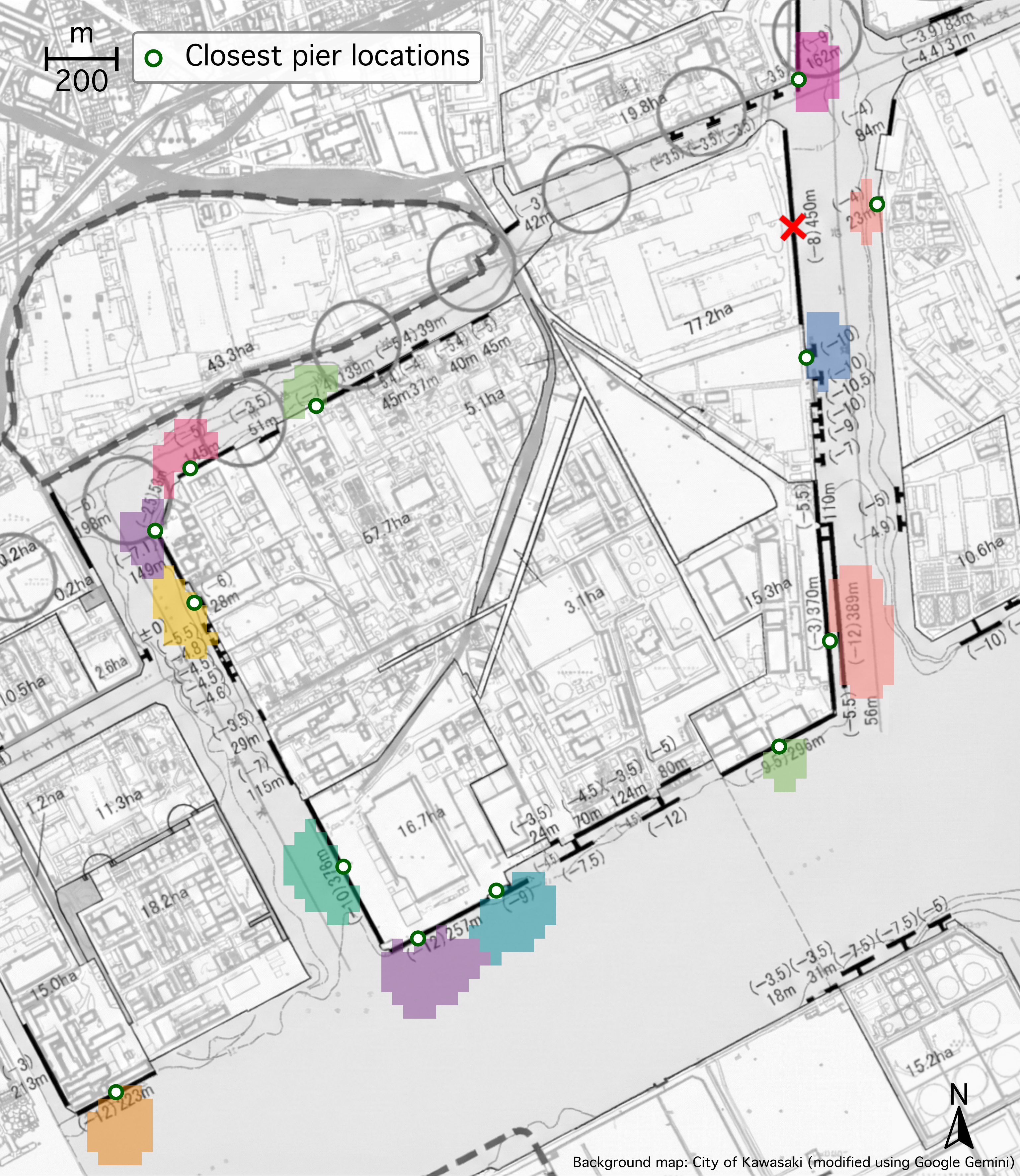}
    };
}
}
}
    \caption{
(\textbf{a}) Average 
 vessel course direction within each grid cell and berth locations (\SI{200}{m} from the shore). Bearings are shaded by the resultant length $\overline{R}$ \citep{Mardia1999} of cell values and the moving-vessel density. The busiest berths (Table~\ref{tab:most_busy_berths}) are highlighted in green, and those with the highest {vessel density} (Table~\ref{tab:most_dense_berths}) in yellow.  The Yokohama-Kawasaki port area is enlarged in Figure~\ref{fig:closeup_kws_yok}b, showing the remaining berth IDs from \mbox{Tables~\ref{tab:most_busy_berths} and \ref{tab:most_dense_berths}}. (\textbf{b}) Close-up of 13 berths identified around Ogimachi Island within Kawasaki Port. The berth areas are shown as colorized pixels at the $\SI{25}{\meter}\times\SI{31}{\meter}$ analysis resolution. Dots indicate the corresponding identified pier locations. The port map in the background is kindly provided by the City of Kawasaki (text labels removed) 
 \cite{KawasakiCity2025}. The red cross is from the original map and denotes planned pier removal, large circles indicate planned road construction.  }

\begin{tikzpicture}[remember picture, overlay]

\def\yOffset{1.806} 

\begin{scope}[yshift=\yOffset cm]
\def\xStart{1.69}
\def\xEnd{8.177}
\def\yStart{12.84} 
\def\yEnd{7.97} 

\def\yStarttwo{13.258}
\def\yEndtwo{16.42}

\def\gapStart{0.496}
\def\gapEnd{0.630}

\pgfmathsetmacro{\xGapStart}{\xStart + \gapStart*(\xEnd-\xStart)}
\pgfmathsetmacro{\yGapStart}{\yStart + \gapStart*(\yEnd-\yStart)}
\pgfmathsetmacro{\xGapEnd}{\xStart + \gapEnd*(\xEnd-\xStart)}
\pgfmathsetmacro{\yGapEnd}{\yStart + \gapEnd*(\yEnd-\yStart)}

\draw[very thin, black] (\xStart,\yStart) -- (\xGapStart,\yGapStart);
\draw[very thin, black] (\xGapEnd,\yGapEnd) -- (\xEnd,\yEnd);
\draw[very thin, black] (\xStart,\yStarttwo) -- (\xEnd,\yEndtwo);
\end{scope}

\end{tikzpicture}
\label{fig:bearing_closeup_maps_tokyo}
\end{figure}
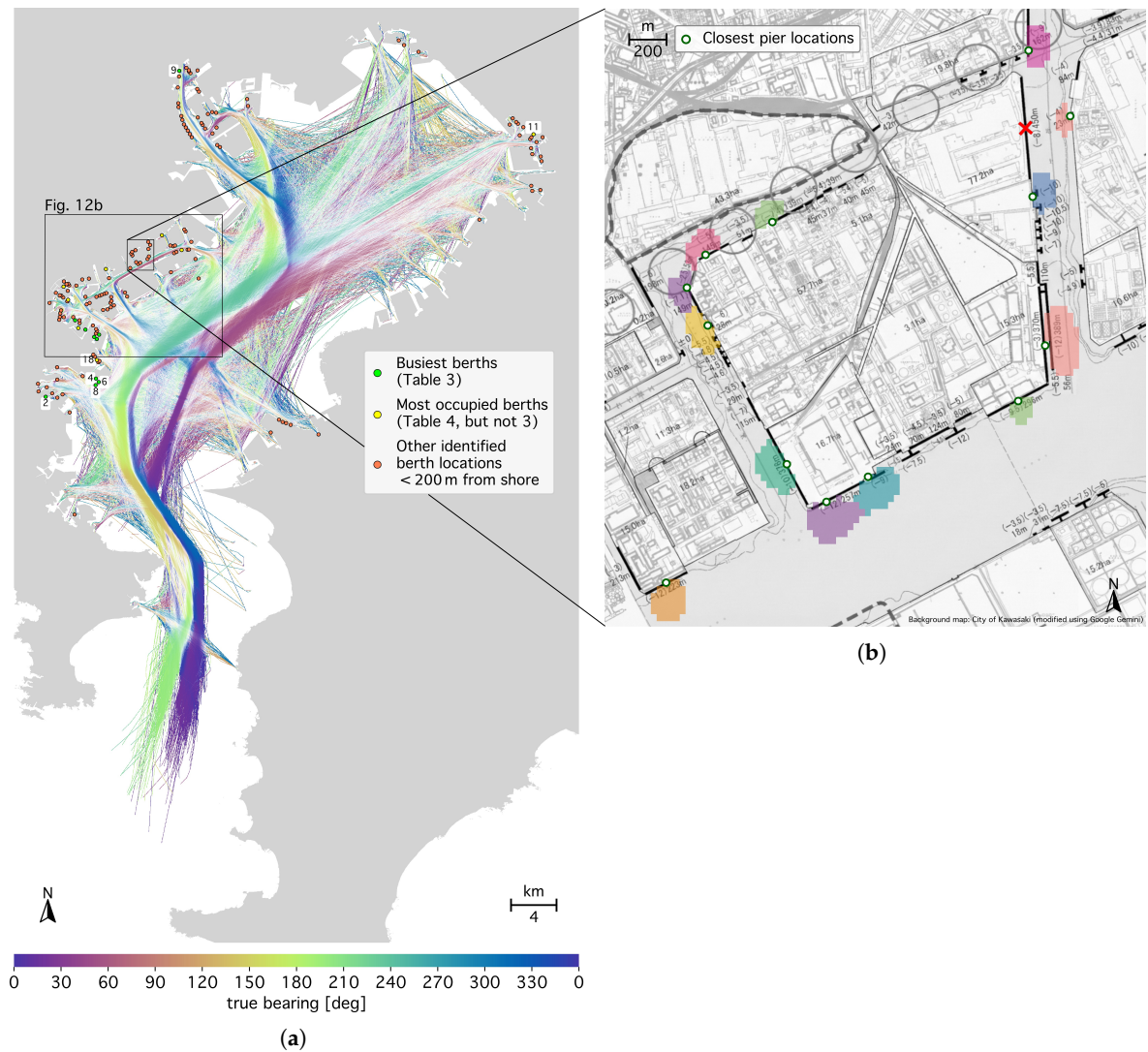

\begin{figure}[ht!] 
\captionsetup[subfloat]{margin=2pt}
    \subfloat[
    \label{fig:density_map_tokyo_closeup_kws_yok}]{\includegraphics[width=0.48015\textwidth]{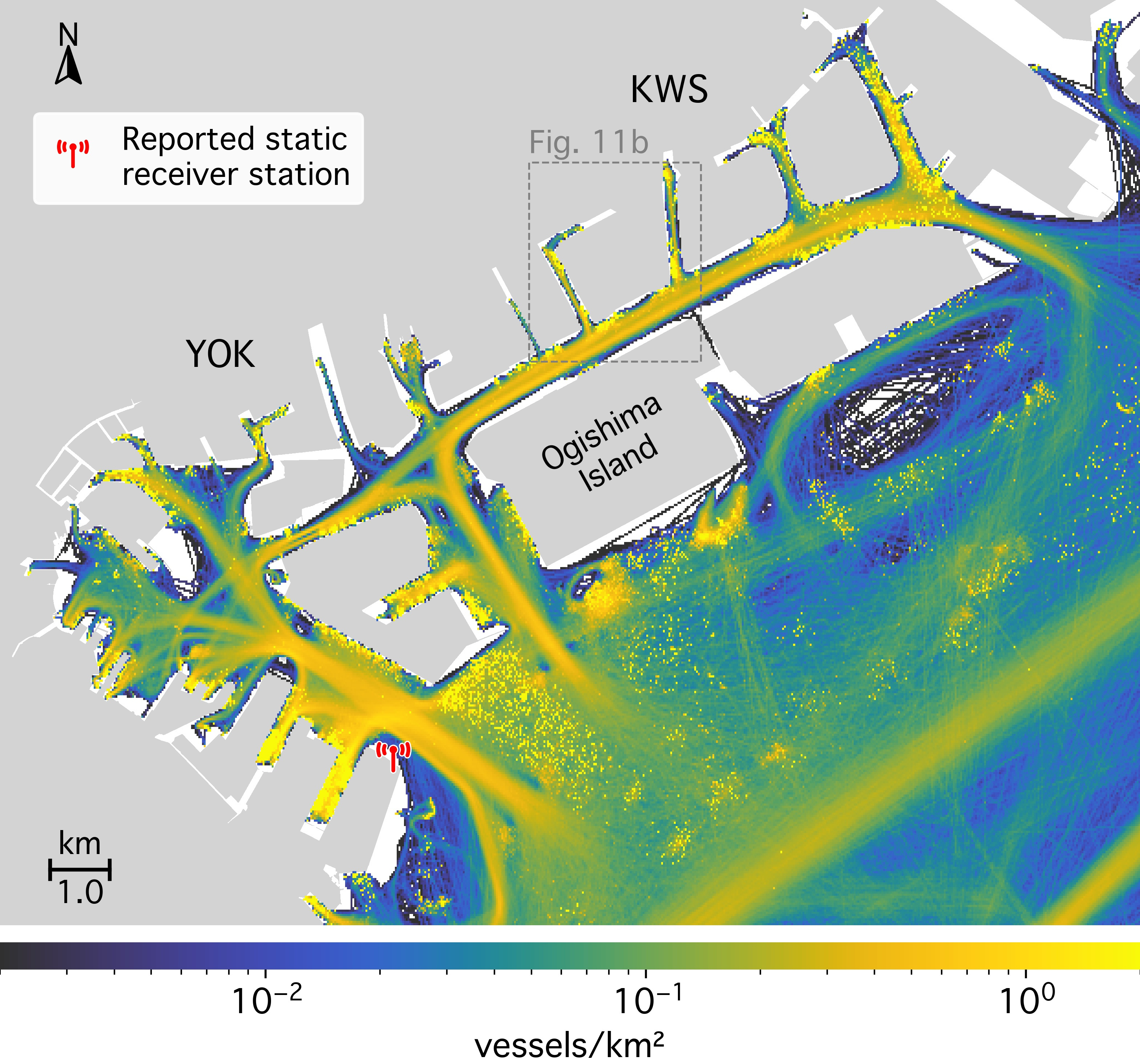}
    }
    \hfill
\subfloat[
\label{fig:bearing_map_tokyo_closeup_kws_yok}
]{
    \includegraphics[width=0.48985\textwidth]{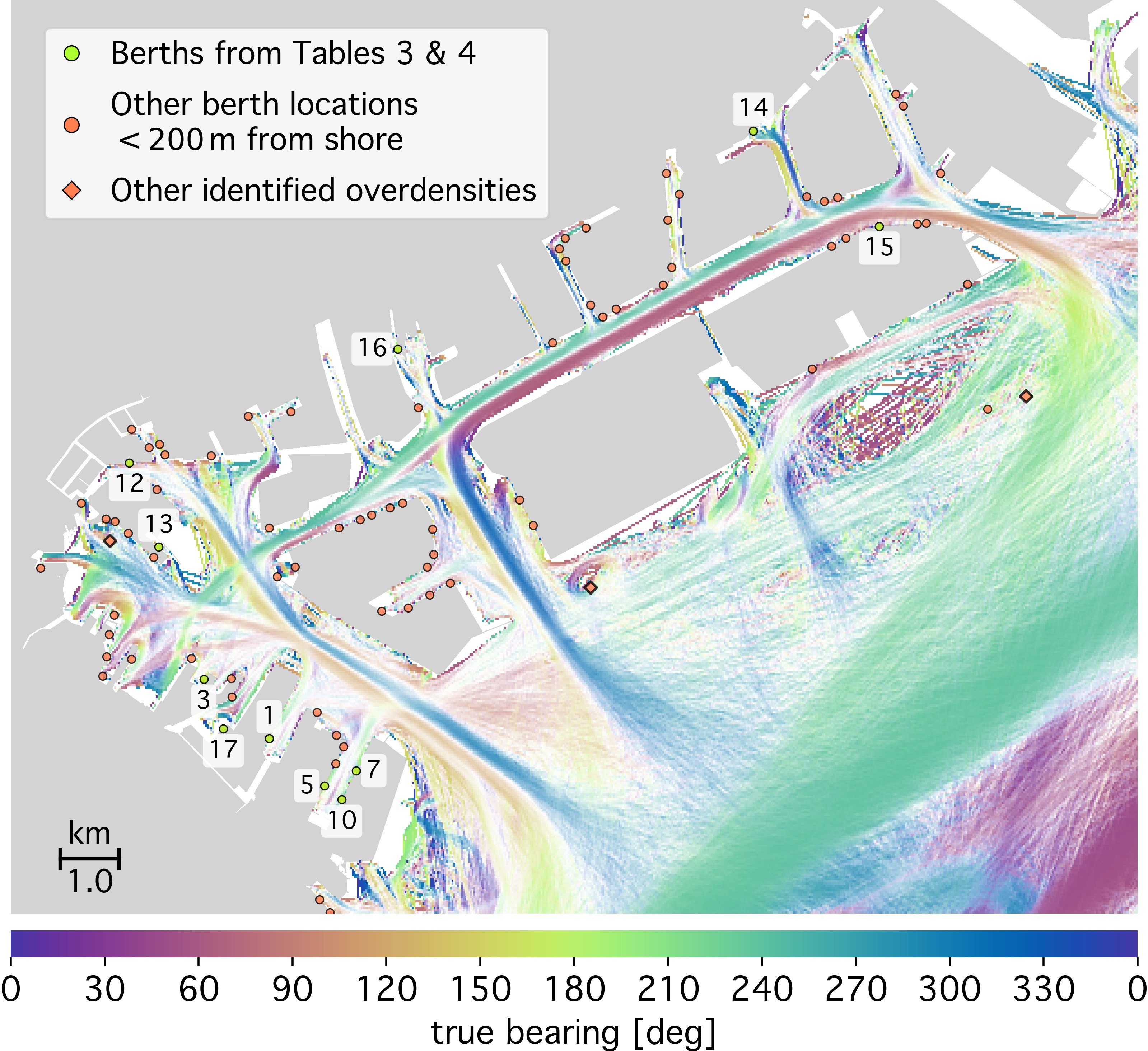}
 }
\caption{
(\textbf{a}) Vessel 
 density in the Yokohama (YOK) and Kawasaki (KWS) port areas. The Yokohama Port Symbol Tower AIS receiver (red tower symbol) and a close-up of the Kawasaki port in Figure~\ref{fig:bearing_closeup_maps_tokyo}b are also marked.
(\textbf{b}) Corresponding vessel movement directions (colormap) and berth locations (dots). Yellow-green dots mark berths with their IDs (ranked by arrival rate) listed in Tables~\ref{tab:most_busy_berths} and \ref{tab:most_dense_berths}. Other overdensities (>\SI{200}{\meter} from the shore) are marked with red diamonds.
     }
    \label{fig:closeup_kws_yok}
\end{figure}

{

In Figures \ref{fig:density_speed_maps_tokyo}--\ref{fig:closeup_kws_yok}, we present the spatial mapping of vessel density and traffic in Tokyo Bay. These figures represent 91\% of vessel movements, excluding erroneous tracklets over land, and 52\% of stationary vessels with mooring positions identified within \SI{500}{\meter}. 

Figure~\ref{fig:density_speed_maps_tokyo}a shows the average total density of moving and stationary vessels. Fairways, traffic separation schemes, prominent anchorages, berthing areas, and fishing grounds in the outer bay are well resolved. The figure also exhibits artifacts due to sparse data, particularly in the outer bay (bottom part of the figure), where vessel routes are not properly reconstructed along traffic lanes. For the Tokyo Bay Ferry crossing the outer bay, signal occlusion by the hilly coast of Cape Kannon falsely classifies ferry boats as moored instead of completing their trips to and from Kurihama on the western side. This effect is also seen on the eastern side and in previous work using data from a different provider \cite{Grinyak2023}.

\begin{figure}[ht!] 
    \centering
    \subfloat[
    \label{fig:density_map_tokyo}]{
    \includegraphics[width=0.485\textwidth, trim=0px 0px 0px 15px, clip]{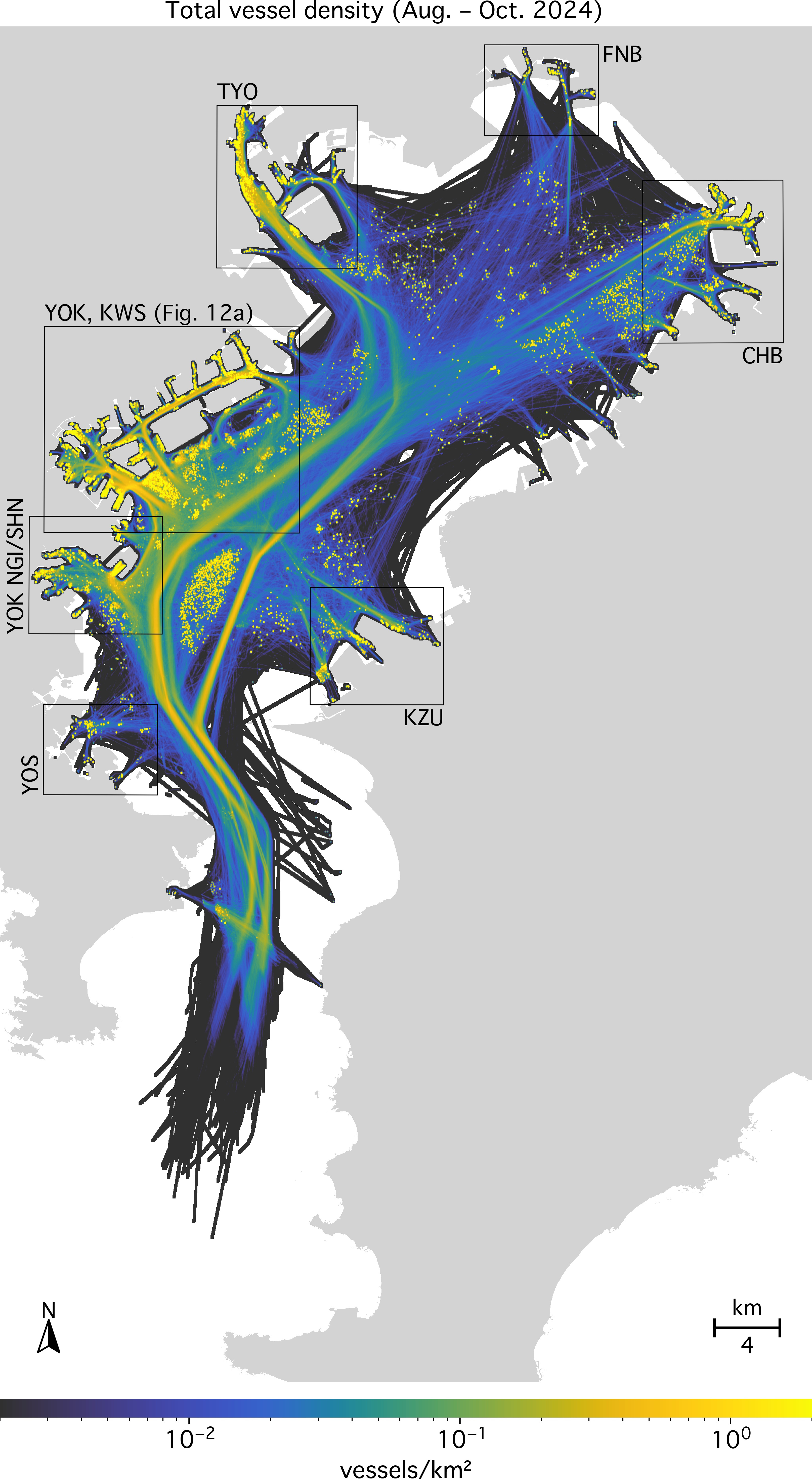}}
    \hfill
        \subfloat[
        \label{fig:speed_map_tokyo}]{
    \includegraphics[width=0.485\textwidth, trim=0px 0px 0px 15px, clip]{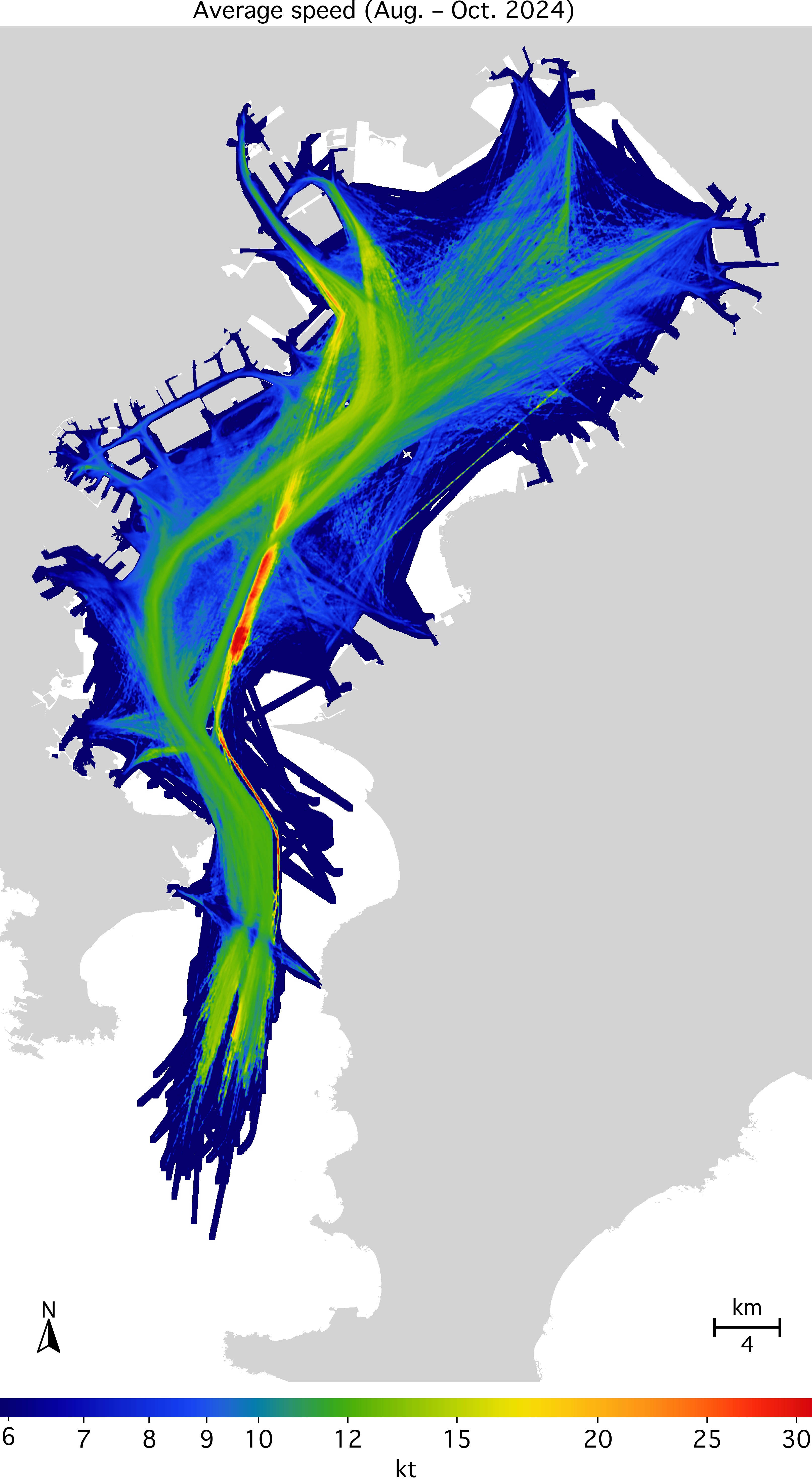}
    }
    \caption{
       (\textbf{a}) Spatially 
 resolved average total vessel density in Tokyo Bay between August and October 2024 (55.1\% of 381 simultaneously active vessels shown). For illustration, black boxes show the major ports within the bay: Tokyo (TYO), Funabashi/Katsunan (FNB), Chiba (CHB), Kisarazu (KZU), Yokosuka (YOS), and Yokohama Negishi and Southern Honmoku (YOK NGI/SHN), as well as the joint Yokohama (YOK) and Kawasaki (KWS) port area, enlarged in Figure~\ref{fig:closeup_kws_yok}a.    (\textbf{b})  Speed averages of moving vessel within each $\sim\,$\SI{30}{\meter} grid cell. For visualization, values are smoothed over 1.5 cells. 
    }
    \label{fig:density_speed_maps_tokyo}
    \end{figure}
}

Figure \ref{fig:density_speed_maps_tokyo}b illustrates the average speed of moving vessels observed in each $\sim\,$\SI{30}{\meter} grid cell. The speed limit of \SI{12}{\knot} in the fairways of the Uraga Channel and around the Nakanose anchorage is largely adhered to, while vessels travel faster in front of Tokyo Port (top part of the figure), as reported by \citet{Shimizu2020}. Moreover, the figure reveals the separate traffic lane for the high-speed hydrofoil ferries operating between Tokyo Port and the Izu Islands. While the track is also visible in the outer bay, in most cases, these ferries make their last AIS contact when leaving and their first contact when entering south of Yokohama in the inner bay, indicated by the bright red area on the high-speed track. Figure \ref{fig:bearing_closeup_maps_tokyo}a shows the course directions of moving vessels, further emphasizing the traffic separation schemes in Tokyo Bay. For example, this figure illustrates that vessels travel only northward through the Nakanose Traffic Route east of Nakanose anchorage in the middle of the bay~\citep{JCG2023}.

Based on the density map segmentation (Section~\ref{sec:vessel_metrics}), we identified 193 berths and anchorage locations in Tokyo Bay with median, mean, and maximum areas of 2.2, 2.7, and 14 hectares. Of these, 161 berth areas (83\%) were within \SI{200}{\meter} of the shore, as displayed in Figure~\ref{fig:bearing_closeup_maps_tokyo}a. Tables \ref{tab:most_busy_berths} and \ref{tab:most_dense_berths} list the ten berths with the highest average arrival frequency and integrated average vessel count. Two of the highest-ranked berths in both categories (JP$\,$YOK$\,$HM$\,$A8 and JP$\,$YOK$\,$HM$\,$YS5; IDs 1 and 3) serve as home berths for tugboats, as does JP$\,$CHB$\,$D$\,$A (ID 11). The remaining berths are primarily used for commercial purposes, except for JP$\,$YOK$\,$Y (ID 17), which harbors pilot boats, and JP$\,$TYO$\,$H$\,$K (ID 9), a passenger pier. As indicated by the percentage ranges in Tables \ref{tab:most_busy_berths} and \ref{tab:most_dense_berths}, relying solely on self-reported vessel type may be unreliable for determining whether a berth is dedicated to a specific operation or serves multiple purposes. Complementary methods, such as usage of SAR imagery \cite{Zhang2025}, may be needed for accurate identification. The berth and pier codes quoted in Tables \ref{tab:most_busy_berths} and \ref{tab:most_dense_berths} were obtained from the \citet{JCG2023}, \citet{YokohamaCity2025}, and \citet{KawasakiCity2025}. The codes closely match the most frequently reported AIS destination codes of arriving vessels, with minor corrections marked in gray in the tables.

Figure \ref{fig:bearing_closeup_maps_tokyo}b provides a further close-up of Ogimachi Island within Kawasaki Port, comparing the 13 identified berths in this area with the port map from the \citet{KawasakiCity2025}. Berth area sizes in this image vary between one and four hectares (\SI{0.01}{\square\km} to \SI{0.04}{\square\km}). Dots indicate the nearest shoreline points to the vessel density centers of mass within these areas and align with the pier locations marked on the map. The visible pixelation corresponds to the $\SI{25}{\meter}\times\SI{31}{\meter}$ map cell size.

Figure \ref{fig:closeup_kws_yok} provides a close-up view of the Yokohama-Kawasaki port area, with busy or occupied  berths  labeled in Figure~\ref{fig:closeup_kws_yok}b by their IDs from \Cref{tab:most_busy_berths,tab:most_dense_berths}.  
The figures provide a detailed view of port activity and traffic flow. For example, vessels larger than 1000 GT enter the Keihin Canal in the center of Kawasaki Port through the Tsurumi Passage in the southwest and leave through the Kawasaki Passage in the northeast \citep{JCG2023}. The figures also highlight the activity at the four liquefied natural gas (LNG) and crude oil sea berths near Ogishima Island (Figure~\ref{fig:closeup_kws_yok}a \citep{JCG2017}). One of these berths is flagged as land in the  \citet{Openstreetmap2024} data and therefore marked with a red dot in Figure~\ref{fig:closeup_kws_yok}b, while the other two appear as offshore overdensities (red diamonds in Figure~\ref{fig:closeup_kws_yok}b). Also, we identify the LNG terminal at Kisarazu port, but not the terminals at Anegasaki port (Figure~\ref{fig:bearing_closeup_maps_tokyo}a).

\begin{table}[ht!]
\caption{Busiest berths in Tokyo Bay (August--October 2024). $\varphi$ and $\lambda$ denote the latitude and longitude of the closest pier. For brevity, the JP prefix is dropped from the port codes transmitted by the vessels, and XX 
from YOK XX$\ldots$ codes in the convention from \cite{JCG2023}. Manual corrections are marked in gray. For all berths, except for the first three, tugboats are excluded from the dominant vessel-type percentages, with the ranges indicating that for some arrivals, the vessel type could not be determined. ``Vessels in berth'' indicates the number of simultaneously moored vessels. The 
supplemental material (see Data Availability Statement at the end of this article)  provides additional information, such as the absolute numbers of arrivals per vessel type.
}
\small
\renewcommand{\arraystretch}{1.2}
\begin{tabular*}{\textwidth}{@{\extracolsep\fill}rclcccc@{}}
\toprule
    \multirow{2}{0.3cm}{\textbf{ID}} & 
   \boldmath$\varphi$ \textbf{[}\boldmath$^\circ$\textbf{]}& 
   \multirow{2}{2.cm}{\textbf{Port/Berth Code}} & 
   \textbf{Dominant} & 
  \textbf{Arrivals} & 
  \textbf{Vessels} &
  \textbf{Area}\\[-0.1cm]
    & 
    \boldmath$\lambda\,$ \textbf{[}\boldmath$^\circ$\textbf{]} & 
    & 
    \textbf{Vessel Type} & 
  \textbf{per Day}  & 
  \textbf{in Berth} &
 \textbf{ [km\boldmath$^2$]}\\[-0.025cm]
\midrule
 1  &
 ${}^{\;\;35.4418}_{139.6695}$ &
 YOK HM \textcolor{gray}{A8}& 
 Tug {(84--99\%)} &
 $30.4$ &
$15.9$ &
$0.14$
 \\
 2 &
 ${}^{\;\;35.3878}_{139.6312}$ &
 YOK  NGI& 
 Tug {(99--100\%)}&
 $12.8$ &
 $\;\,0.3$ &
 $0.02$
\\
 3 &
 ${}^{\;\;35.4492}_{139.6596}$ &
 YOK HM \textcolor{gray}{YS5} &
 Tug {(86--100\%)} &
 $12.4$ &
 $\;\,7.7$ &
 $0.09$
\\
 4 &
 ${}^{\;\;35.4026}_{139.6798}$ &
 YOK SHN MC2 &
 Container {(57--96\%)}&
 $\;\,9.9$ &
 $\;\,0.9$ &
 $0.06$
 \\
  5  &
 ${}^{\;\;35.4360}_{139.6781}$ &
 YOK HM C7 & 
 Container  {(35--78\%)}&
 $\;\,8.5$ &
 $\;\,1.2$ &
 $0.07$
 \\
  6  &
 ${}^{\;\;35.3998}_{139.6826}$ &
 YOK \textcolor{gray}{S}HN \textcolor{gray}{MC3} & 
 Container  {(50--94\%)}&
 $\;\,8.5$ &
 $\;\,0.8$ &
 $0.06$
 \\
  7  &
 ${}^{\;\;35.4378}_{139.6828}$ &
 YOK HM \textcolor{gray}{D}4 &  
{Container (18--90\%)} &
 $\;\,7.9$&
 $\;\,0.7$ &
 $0.04$
 \\
  8  &
 ${}^{\;\;35.3972}_{139.6796}$ & 
 YOK SHN \textcolor{gray}{MC4} & 
 Container {(48--96\%)}&
 $\;\,7.3$&
 $\;\,0.8$ &
 $0.06$
 \\
  9  &
 ${}^{\;\;35.6496}_{139.7603}$ &
 TYO H \textcolor{gray}{K} &  
 Passenger  {(100\%)}&
 $\;\,4.6$ &
 $\;\,1.1$ &
 $0.04$
 \\
   10  &
 ${}^{\;\;35.4342}_{139.6807}$ &
 YOK HM D1& 
{Container (42--96\%)} &
 $\;\,4.5$ &
 $\;\,0.5$ &
 $0.04$
 \\
\bottomrule
\end{tabular*}
\label{tab:most_busy_berths}
\end{table}\vspace{-9pt}

\begin{table}[ht!]
\caption{Most occupied berths in Tokyo Bay (August--October 2024). The name of berth ID 13 is abbreviated from ``YOKHAMA NORTH DOCK'' (sic). All columns are defined the same as in Table~\ref{tab:most_busy_berths}.}
\renewcommand{\arraystretch}{1.3}
\small
\begin{tabular*}{\textwidth}{@{\extracolsep\fill}rclcccc@{}}
\toprule
    \multirow{2}{0.3cm}{\textbf{ID}} & 
   \boldmath$\varphi$ \textbf{[}\boldmath$^\circ$\textbf{]}& 
   \multirow{2}{2.cm}{\textbf{Port/Berth Code}} & 
   \textbf{Dominant} & 
  \textbf{Arrivals} & 
  \textbf{Vessels} &
  \textbf{Area}\\[-0.1cm]
    & 
    \boldmath$\lambda\,$ \textbf{[}\boldmath$^\circ$\textbf{]} & 
    & 
    \textbf{Vessel Type}& 
  \textbf{per Day}  & 
 \textbf{ in Berth} &
 \textbf{ [km\boldmath$^2$]}\\[-0.025cm]
\midrule
 1  &
 ${}^{\;\;35.4418}_{139.6695}$ &
 YOK HM \textcolor{gray}{A8}& 
 Tug {(84--99\%)} &
 $30.4$ &
$15.9$ &
$0.14$
\\
 3  &
 ${}^{\;\;35.4492}_{139.6596}$ &
 YOK HM \textcolor{gray}{YS5} & 
  Tug {(86--100\%)}&
 $12.4$ &
 $\;\,7.7$ &
 $0.09$
\\
 11 &
 ${}^{\;\;35.5984}_{140.1053}$ &
 CHB D A &
 Tanker {(80--92\%)}&
 $\;\,3.7$ &
 $\;\,2.6$ &
 $0.14$
 \\
  12  &
 ${}^{\;\;35.4762}_{139.6481}$ &
 YOK D \textcolor{gray}{B}(HS) & 
 General cargo {(41--68\%)} &
 $\;\,3.9$ &
 $\;\,2.5$ &
 $0.08$
 \\
 13  &
 ${}^{\;\;35.4658}_{139.6526}$ &
 YOK NORTH DOCK$\!\!\!$ & 	
 Military  {(42--100\%)}&
 $\;\,0.1$ &
 $\;\,2.2$ &
 $0.04$
 \\

 14  &
 ${}^{\;\;35.5177}_{139.7436}$ &
 KWS SG &  
 Tanker {(15-99\%)} &
 $\;\,1.9$&
 $\;\,2.1$ &
 $0.07$
 \\
  15  &
 ${}^{\;\;35.5058}_{139.7630}$ &
 KWS HO \textcolor{gray}{26}& 
 Tanker  {(59--78\%)}&
 $\;\,3.5$ &
 $\;\,1.8$ &
 $0.09$
 \\
 16  &
 ${}^{\;\;35.4904}_{139.6892}$ & 
 YOK SH 	 & 	
 General cargo {(12--94\%)} &
 $\;\,0.8$&
 $\;\,1.8$ &
 $0.03$
 \\
 17  &
 ${}^{\;\;35.4430}_{139.6626}$ &
 YOK \textcolor{gray}{Y} & 
 Pilot {(100\%)}&
 $\;\,1.3 	$ &
 $\;\,1.7$ &
 $0.04$
 \\
 18  &
 ${}^{\;\;35.4174}_{139.6807}$ &
 YOK SH\textcolor{gray}{N} & 
 General cargo  {(1--100\%)}&
 $\;\,4.0$ &
 $\;\,1.5$ &
 $0.07$
 \\
\bottomrule
\end{tabular*}
\label{tab:most_dense_berths}
\end{table}

\subsection{Localized Receivers}
\label{sec:results_basestation}

Figure~\ref{fig:inferred_receivers_close}b,c in Section~\ref{sec:methods} display the inferred positions, containment ranges, and confidence ellipses of the Tokyo and Yokohama AIS receiver stations.  

For the  weighted distribution of the Tokyo receiver intersections, we obtained the lengths of the semi-axes $a$, $b$ as \SI{1.51}{\km} and \SI{390}{\meter} in distance units, respectively. Figure~\ref{fig:inferred_receivers_close}b shows that the weighted mean and median focal point positions (blue square and green upright cross) are found within \SI{100}{\meter} and within their statistical confidence ellipses (blue dashed and green dashed-dotted ellipses). The ellipses also align with the unweighted median position (yellow diagonal cross and dotted ellipse). The unweighted mean focal point (red diamond) is located \SI{200}{\meter} away, with a large uncertainty (purple dashed-double-dotted ellipse). The black ellipse outlines the region containing 68\% of all pairwise intersections, covering an area of \SI{4.2}{\square\km}. All positions are located within \SI{400}{\meter} of the AIS receiver station 3752 reported by  AISHub~\cite{AISHub2025} (black dot) and lie within the precision with which AISHub discloses its position (gray-shaded area of \SI{0.9}{\square\km}; all positions and confidence ellipses are numerically listed in Table~\ref{tab:receiver_positions} in the appendix). We conclude that the same station likely contributes to the data stream from \citet{aisstream2025} and that the majority of AIS messages from Tokyo Bay are relayed via this receiver. However, the aisstream.io network does not report any receiver station in this area. 

Very high frequency (VHF) radio waves can propagate unblocked by the horizon, ignoring elevations along the line of sight (decreasing the range) and atmospheric ducting (increasing the range; \citep{Craig2003}), over a maximum geometric distance of
\begin{equation}
   d \lesssim \sqrt{\frac{2\,k\,R_\text{E}}{\SI{1000}{\km}}}\left(\sqrt{\frac{h_\text{R}}{\SI{1}{\meter}}} + \sqrt{\frac{h_\text{T}}{\SI{1}{\meter}}} \right)\,\SI{}{\km}\,,
   \label{eq:radiodistance}
\end{equation}
where the factor $k\approx 4/3$ accounts for atmospheric refraction \citep{Craig2003}, $R_\text{E}$ is the Earth's radius, and $h_\text{T}$ and $h_\text{R}$ are the heights above sea level of the transmitter and receiver, respectively. Assuming that the Tokyo receiver station receives messages from the Tokyo Bay Ferry at $d\approx\SI{45}{\km}$, as suggested by the radio shadows, with the ferry vessels transmitting at $h_\text{T}\lesssim\SI{20}{\meter}$, the receiver  must be located at $h_\text{R}\gtrsim\SI{40}{\meter}$, or more than about $\SI{20}{\meter}$ above ground within the black ellipse in Figure~\ref{fig:inferred_receivers_close}b \cite{NOAA2022}. According to satellite imagery, no such high buildings or antennas are found within the 95\% statistical confidence cone of the weighted mean or median in Figure~\ref{fig:inferred_receivers_close}b. The closest sufficiently tall structures are four $\sim\,$$\SI{100}{\meter}$-high buildings located at the position of the mean unweighted focal point, the tallest of which has a height of \SI{151}{\meter} above ground \cite{TallBuildings2025}, or \SI{165}{\meter} above sea level (black solid cross in Figure~\ref{fig:inferred_receivers_close}b).
Therefore, while the weighted mean value suggests that we locate the Tokyo receiver position within \SI{0.06}{\square\km} at the 95\% confidence level (\SI{0.1}{\square\km} at 99\%), the uncertainty is probably underestimated due to unaccounted correlation between the radio-shadow fits. Systematic bias may also increase the uncertainty, for example, from fitting radio shadows to AIS message positions binned in a Euclidean plane, or by ignoring diffraction. Moreover, the distribution of intersections has wider tails than a Kent distribution, as shown in Figure~\ref{fig:mysterious_receiver_wide_weighted} in the appendix. 

However, we argue that with three months of AIS data, the receiver’s position can be determined with a precision equal to or better than what is publicly disclosed. This raises concerns about data protection for receivers who wish to remain private. When sharing their data, operators should be aware that their identities may be inferred, even when they function solely as passive radio receivers and do not disclose their positions to the network or its users. We remark that apart from this analysis, in the case of the Tokyo receiver station, the building where the receiver is most likely installed can also be identified based solely on the publicly disclosed location precision and a reasonable height constraint.

The second AIS receiver in the Yokohama area (Figure~\ref{fig:inferred_receivers_close}c) is inferred with lower precision, with the mean and median positions differing by several kilometers. 68\% of intersections fall in an area of \SI{5.9}{\square\km} (black ellipse in Figure~\ref{fig:inferred_receivers_close}c), and the 95\% weighted mean confidence ellipse spans over \SI{4.5}{\square\km}. Two candidate receiver stations are found in the vicinity: the Yokohama Port Symbol Tower (approximately \SI{60}{\meter} above sea level) reported by the \citet{aisstream2025}  network (red tower symbol in Figure~\ref{fig:inferred_receivers_close}c) and the amateur radio unit JR1CAD (\SI{140}{\meter} above sea level) contributing as station 5552 to the network from VesselFinder~\cite{VesselFinder2025} (black dot). Because all associated radio shadows are found at very small mutual angles, the weighting may bias the inferred position toward smaller distances from the fitted segments. We therefore conclude that the receiver is likely associated with the more distant station 5552/JR1CAD from VesselFinder. However, ambiguity remains due to the limited data sample, as well as the possibility that a third unknown station may be the data source.

\section{Discussion}
\label{sec:discussion}

Several uncertainties and limitations affect our analysis, which we discuss in detail below. Some of these uncertainties have been quantified and are included in our results presented in Figures~\ref{fig:hist_times_vessels_cart_tokyo}, \ref{fig:vessel_flux_tokyo}, and \ref{fig:vessels_historic_tokyo}, and in \Cref{tab:vessel_numbers,tab:vessel_numbers_flux,tab:most_busy_berths,tab:most_dense_berths}.

\subsection{Uncertainty About Classifying Vessels as Leaving or Moored }
\begin{figure}[ht!]
\includegraphics[width=0.55\textwidth]{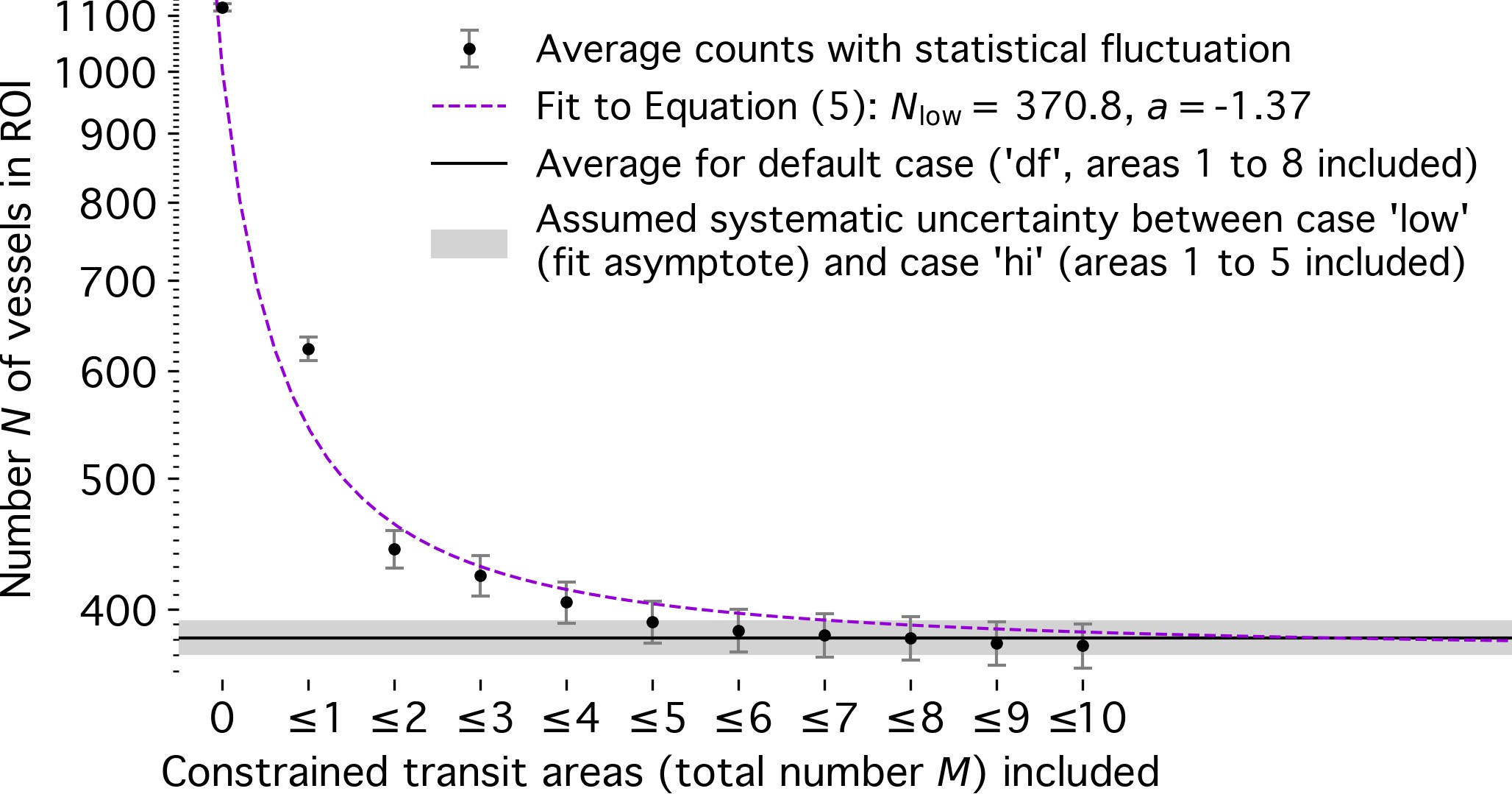}
    \caption{
    Estimated average vessel count $N$ in Tokyo Bay as a function of the cumulative number $M$ of successive transit areas (Figure~\ref{fig:roi_exit_areas_tokyo}), used to assess the uncertainty in the vessel status classification. 
    } 
    \label{fig:vessel_limit_tokyo}
\end{figure}

For the vessel status classification from Section~\ref{sec:data_processing}, we quantified the uncertainty of confusing vessel absence with mooring by varying the size of the cumulative transit area shown in Figure~\ref{fig:roi_exit_areas_tokyo}. By default, the analysis included successive areas 1 to 8, giving vessel numbers $N_\text{df}$ and transit rates $\dot{N}_{\text{df}}$. For a lower bound on the overall transit rate, resulting in an upper estimate for the vessel count, $N_\text{hi}$, we only added successive areas 1 to 5 from Figure~\ref{fig:roi_exit_areas_tokyo} to the main transit area, covering the fairways up to the Tokyo Bay Aqua-Line bridge tunnel. Using all areas in Figure~\ref{fig:roi_exit_areas_tokyo}, including areas 9 and 10, increases the likelihood of false-positive transits. As illustrated in Figure~\ref{fig:lost_signal_example}, signal loss can also occur within the bay when vessels veer away from the strongest receiver station, which we assume to be the Tokyo receiver (red disk; see also Section~\ref{sec:results_basestation}). However, the vessel count converges as the bay is increasingly covered with successive transit areas, as shown in Figure \ref{fig:vessel_limit_tokyo}. This indicates that the constraints from Equation~(\ref{eq:t_thresh}) effectively differentiate between absence and mooring. We estimated a lower limit of the vessel count, $N_\text{low}$, from the convergence trend by fitting the ad hoc function 
\begin{align}
    N= N_{\text{low}}\times \exp\bigl(\left(M+1\right)^{a}\bigr)\,,\quad a<0\,, 
    \label{eq:fit}
\end{align}
to the trend of cumulatively included successive transit areas, $M$, with $N\rightarrow N_{\text{low}}$ for \mbox{$M\rightarrow\infty$}, as displayed in Figure~\ref{fig:vessel_limit_tokyo}. This fit was done under the assumption of an approximately similar size for all areas $M$. To obtain the transit rate $\dot{N}_{\text{low}}$ corresponding to the vessel count $N_{\text{low}}$, we assumed that every vessel discounted from the default case, $N_\text{df}-N_{\text{low}}$, was absent on average but shortly left and re-entered once a day, such that 
\begin{align}
    \dot{N}_{\text{low}}= \dot{N}_{\text{df}} + 2\;\frac{N_\text{df}-N_{\text{low}}}{\SI{1}{\day}}\,.
\end{align}

\Cref{tab:case_comparison} 
 in the appendix details the default, lower, and upper estimates for the vessel counts and corresponding transit rates. We also calculated these estimates separately for each of the six vessel categories. However, we assumed a global uncertainty $\delta_\text{low} =2.7\%$ instead of applying the fit to Equation~(\ref{eq:fit}) individually. For the overall vessel activity, the uncertainty from the absence-mooring confusion remained within 5\% for the all-vessel count within the bay and within 10\% for the all-vessel transit rate (Table~\ref{tab:case_comparison}). For the vessel count, it also remained within 6\% for each category; however, the transit-rate uncertainty grew as large as 41\% for pilot, tug, and construction vessels.

We also tested the robustness of the method to ensure that moored vessels were not falsely classified as absent. To this end, we used the successive transit areas 2 to 4, which cut through the Nakanose anchorage in the middle of the bay (under the label `3' in Figure~\ref{fig:roi_exit_areas_tokyo}). No significant discontinuity was observed at the boundary between areas 2 and 3 (or 3 and 4) when only successive areas 1 to 2 (or 1 to 3) were included in the vessel status classification. We therefore conclude that Equation~(\ref{eq:t_thresh}) with $t_0=\SI{48}{\hour}$ does not produce a significant number of false-positive transits.

\subsection{Uncertainty About Vessels Not Using the AIS}

\begin{figure}[ht!] 
        
    \subfloat[    
\label{fig:ratio_of_all_tracked_vessels_tokyo}
    ]{\includegraphics[width=0.23\textwidth]{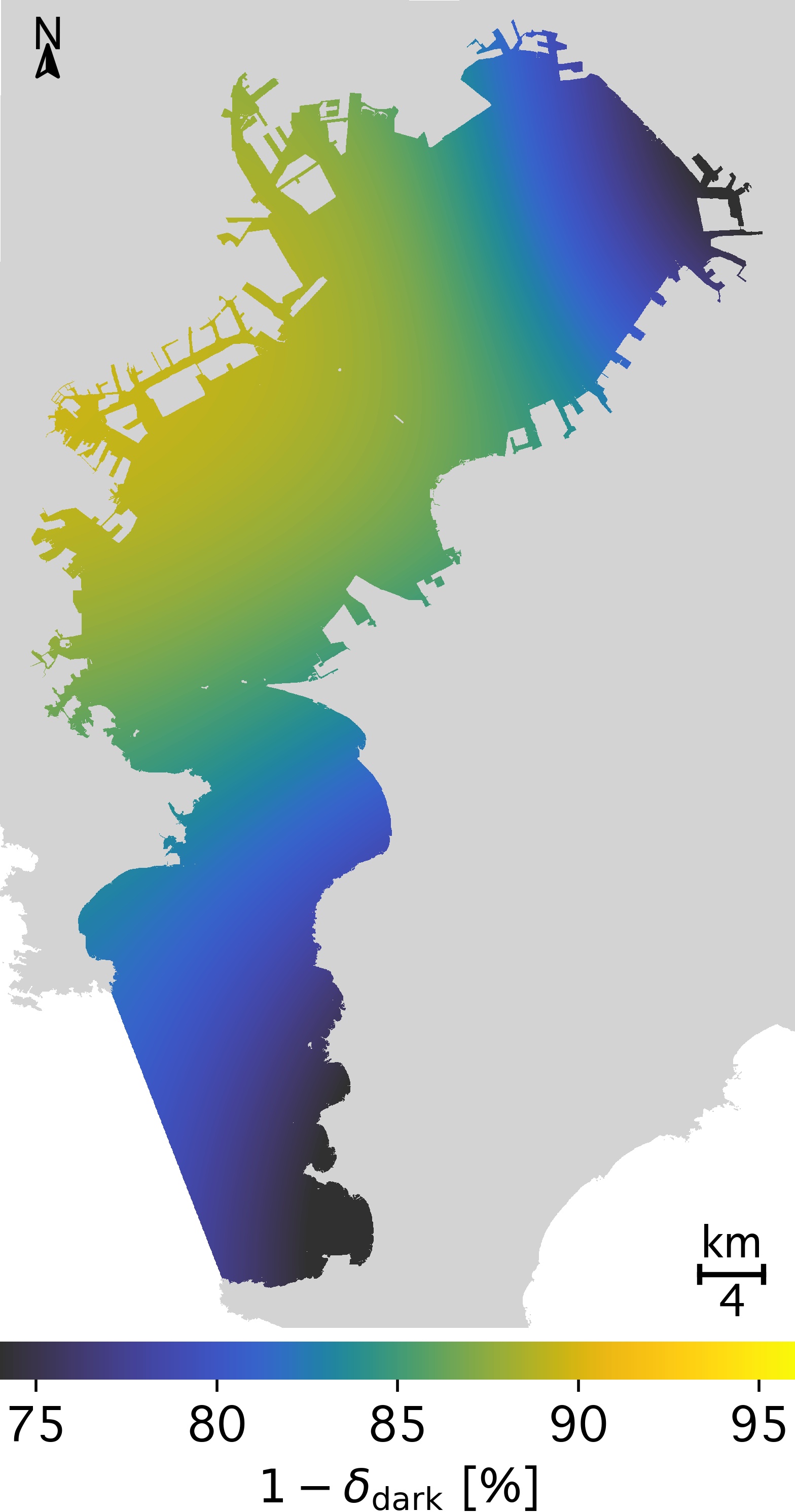}}
    \hfill    
    \subfloat[  
   \label{fig:ratio_of_fishing_vessels_among_all_vessels_tokyo}
    ]{\includegraphics[width=0.23\textwidth]{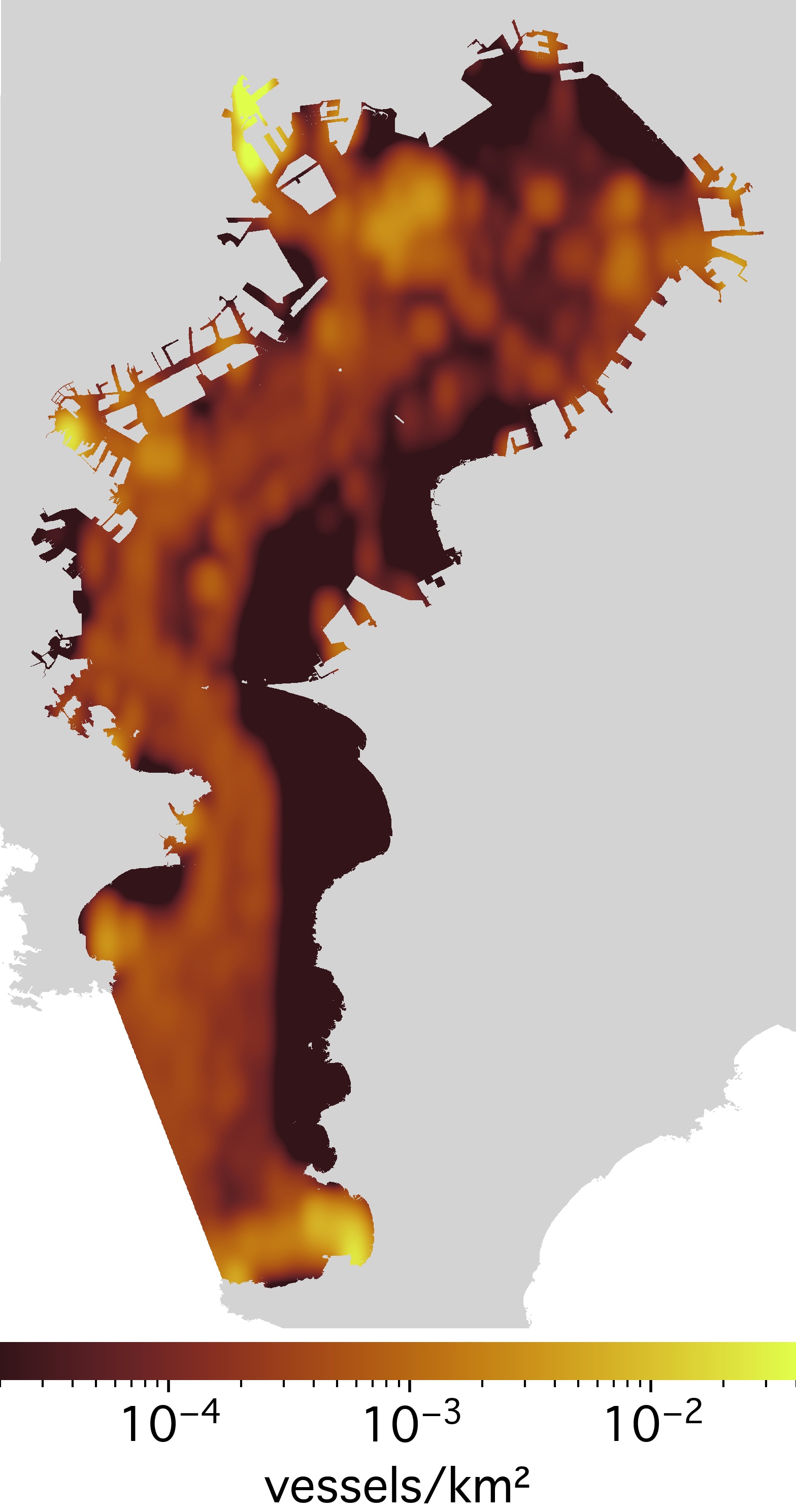}}
	\hfill
    \subfloat[   
\label{fig:ratio_of_tracked_non_fishing_vessels_tokyo}
    ]{\includegraphics[width=0.23\textwidth]{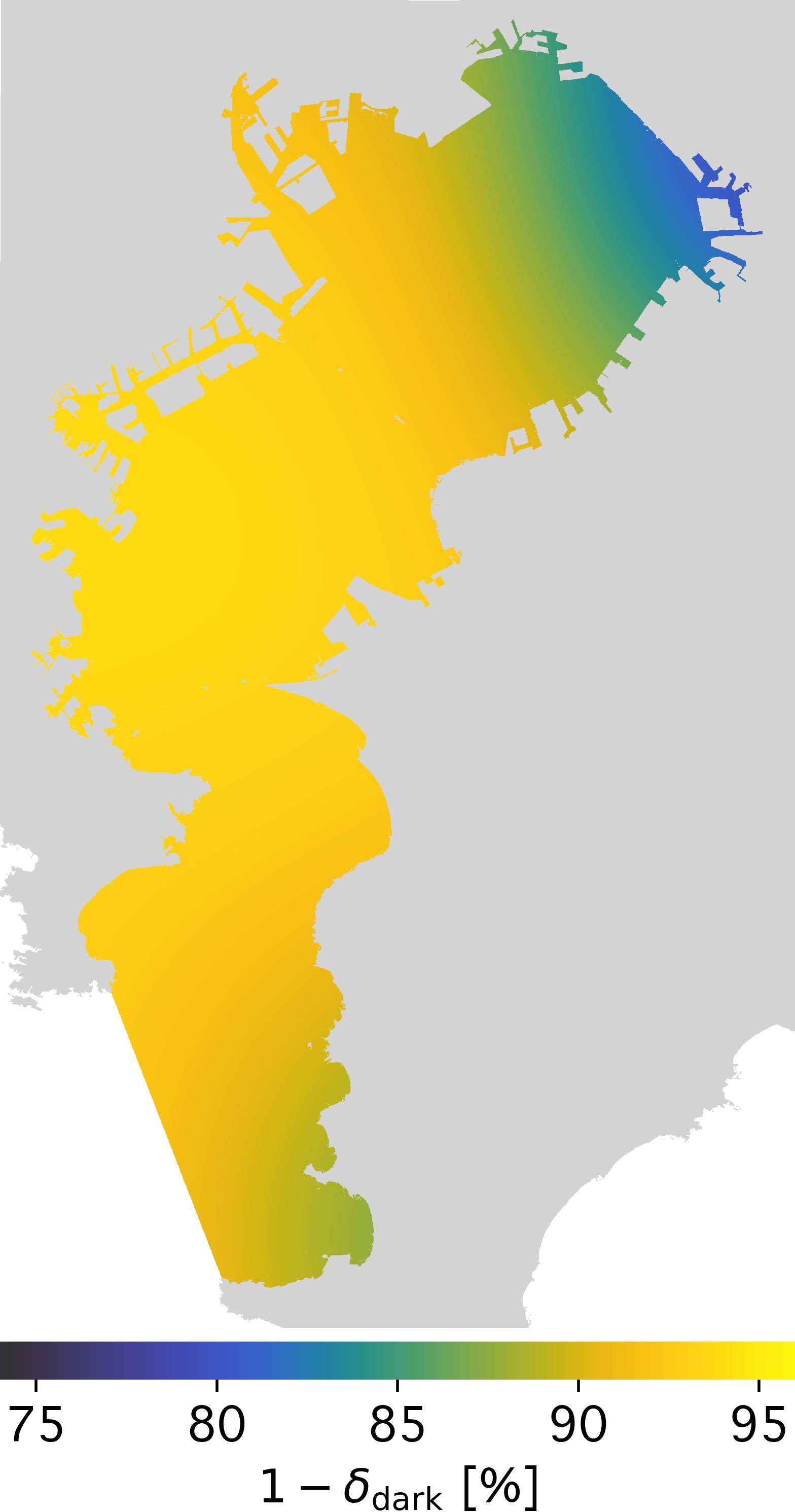}}
    \hfill    
    \subfloat[   
\label{fig:ratio_of_tracked_fishing_vessels_tokyo}
    ]{\includegraphics[width=0.23\textwidth]{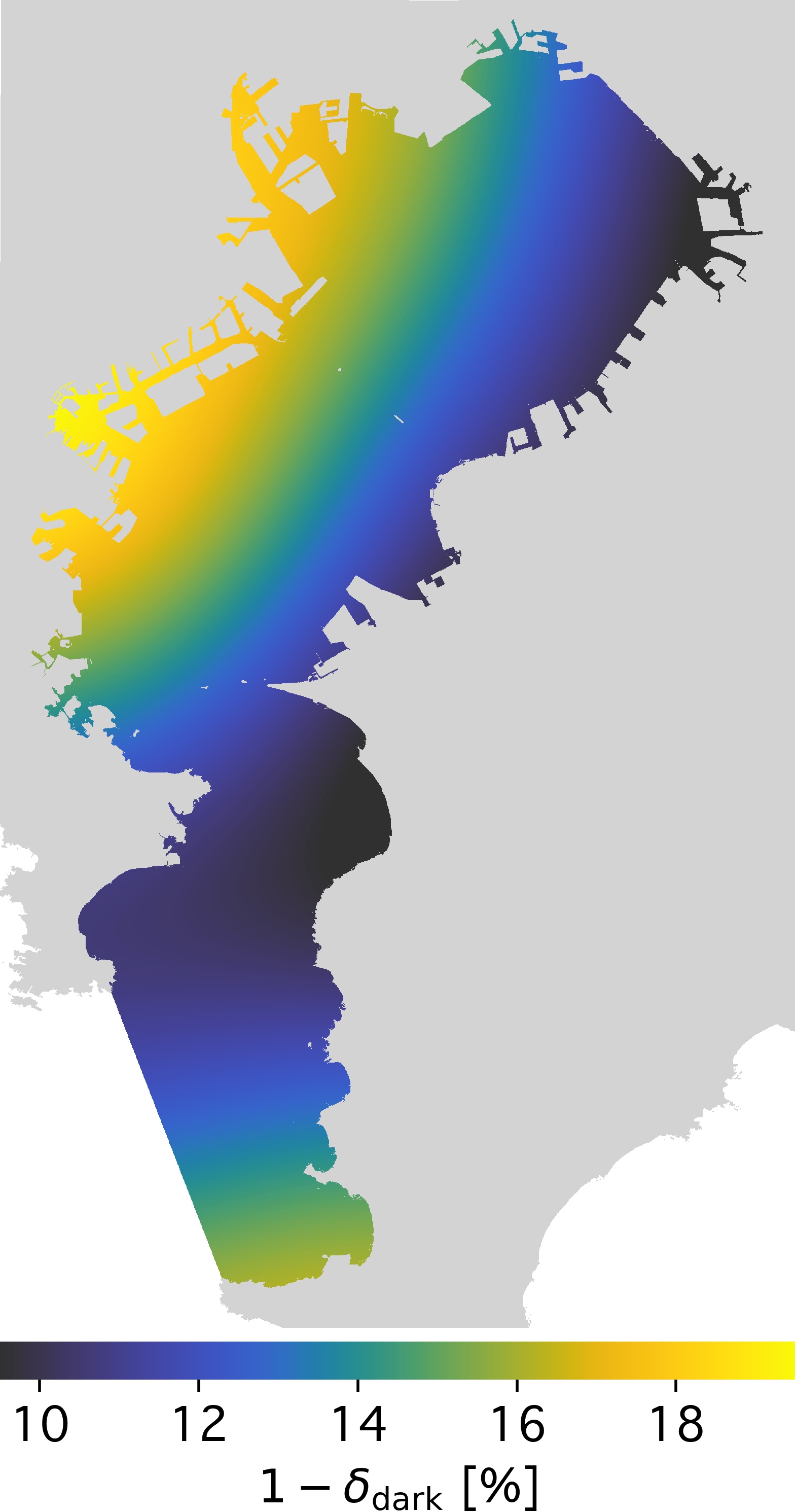}}
        \caption{
(\textbf{a}) Fraction of  vessels tracked by the AIS among all vessels. 
(\textbf{b}) Fishing vessel density by AIS data.
(\textbf{c}) Fraction of AIS-tracked vessels among non-fishing vessels.
(\textbf{d}) Fraction of AIS-tracked vessels among fishing vessels.
  Data shown in panels (\textbf{a},\textbf{c},\textbf{d}) is taken from \citet{Paolo2024} (2017 to 2021), and data shown in panel (\textbf{b}) is taken from \citet{Kroodsma2018} (2012 to 2016 average). 
    } 
    \label{fig:tracked_vessels}
\end{figure}

We used data from \citet{Paolo2024} to estimate the number of vessels not using the AIS. Paolo et al. provide binned data at a resolution of $(\Delta\varphi,\,\Delta\lambda)=(0.2^\circ,\,0.2^\circ)$, with $\sim\,$7~data points in the Tokyo Bay area. We extrapolated these data using a rolling average and the Gauss--Seidel method, and interpolated with cubic splines. Figure \ref{fig:tracked_vessels}a,c,d show the expanded data in Tokyo Bay. On average, for all vessels within the ROI (Figure~\ref{fig:tracked_vessels}a), we obtained $\delta_\text{dark}=16\%$ untracked vessels in the bay. Although this value was lower for non-fishing vessels (9\%, Figure~\ref{fig:tracked_vessels}c) and much larger for fishing vessels alone (87\%, Figure~\ref{fig:tracked_vessels}d), we adopted a global value for simplicity to quantify the underestimation of vessel activity.
Figure~\ref{fig:tracked_vessels}b illustrates the AIS-tracked fishing activity  at a higher resolution.\vspace{-9pt} 

\subsection{Uncertainty About AIS-B Vessels}

We did not include vessels using AIS-B in the main analysis. However, we performed a coarse analysis over a full year of data and found that the ratio of AIS-B vessels in the data was stable throughout 2024, with a monthly average between 11 and 13\%. Table \ref{tab:vessel_aisb} in the appendix provides an approximation of these vessels per category, based on the total observed vessels in one month of data. Support, construction, and other small vessels are primarily affected by this uncertainty. Scaling these per-category percentages onto the observed transit rates (Table~\ref{tab:vessel_numbers_flux}) showed that AIS-B vessels constitute 7\% of those moving into or out of the bay, somewhat less than the proportion of AIS-B vessels present within the bay.

\subsection{Overall Quantification of Vessel-Count and Transit-Rate Uncertainties}

We combined all quantified uncertainties (status confusion, dark, and AIS-B vessels) in quadrature, following Equation~4 from \cite{Huetten2025a}.  Dark and AIS-B vessels caused activity underestimation and are therefore only reflected in the upper uncertainty bound. Since the transit uncertainty extended across most of the Bay Area, it also accounted for the uncertainty in stationary vessels within the Bay. The status confusion did not affect the moving-vessel count because the status classification only affected absences and stationary periods, 
not movements. 
Following \cite{Huetten2025a}, we conclude, based on the 1.5\% of discarded messages (see Appendix~\ref{app:data_processing1}), that neglecting simultaneously active vessels sharing the same MMSI is negligible compared to other uncertainties.

\subsection{Unquantified Limitations of This Study}

In this work, we analyzed only a three-month interval and, when comparing our results with annual averages, assumed that vessel activity is stable throughout the year. Although Figure~\ref{fig:ais_signals_time_tokyo} indicates no long-term variation, and a coarse analysis of received message rates over the year supports this assumption, vessel activity may still fluctuate due to economic factors, as well as the rainy and typhoon seasons. While we expect such fluctuations to be minor, vessel activity is clearly not constant on multi-year timescales, as illustrated in Figure~\ref{fig:vessels_historic_tokyo}. Therefore, future work should address a consistent analysis on annual and multi-year scales.

Furthermore, although we quantified activity underestimation due to unaccounted vessels, future analyses should also include AIS-B vessels, even though AIS-B usage is voluntary and less reliable. To further address missing information on small and dark vessel activity, SAR observations could provide valuable information to complement the AIS data \cite{Paolo2024,Li2025}. In addition, cellphone data may offer a new means of identifying vessel activity in coastal areas, as recently demonstrated for Tokyo Bay in \cite{Imai2025}.


\section{Conclusions}
\label{sec:summary}
Using open-access AIS data from August to October 2024, we reconstructed vessel activity in Tokyo Bay at 30 m spatial resolution and about one-minute temporal precision, deriving average vessel counts and traffic rates, and identifying the most frequented berths and anchorages in the bay. We accounted for uneven and fading receiver coverage and quantified uncertainties from missing data. Within these uncertainties, we found an accelerating shift toward fewer but larger vessels in commercial shipping, mirroring a trend seen globally \citep{UNCTAD2024a} and in other regions of the world \citep{Huetten2025a}.

Our results align with official statistics of maritime activity in Tokyo Bay. We found that open-access AIS data, partly collected by amateur radio operators, can provide insights comparable to those from governmental and proprietary sources, enabling high-resolution and real-time analysis of vessel activities in port areas and congested maritime zones. The presented algorithm is robust across spatial scales, from coastal macroregions to  port dimensions, and incorporates a density-thresholding approach for port, anchorage, or berth detection that complements previous work on this topic (e.g., \citep{Yan2022,AsianDevelopmentBank2023,Zhang2023b,Iphar2024,Qiang2025,Hadjipieris2025}). This may support the shipping industry and port operators in strategic, tactical, and operational planning and contribute to enhanced maritime safety. In addition, it enables independent monitoring of commercial practices and macroeconomic policies. For example, long-term observation of vessel activity in major industrial areas such as Tokyo Bay may provide insights into economic transition processes and 
into alignments with corporate and national greenhouse gas reduction targets \citep{Aslanoglu2024}.

Additionally, we demonstrated that in highly urbanized areas, vessel AIS data can reveal receiver locations with a precision of several \SI{100}{\meter} or better. This may raise concerns for operators of terrestrial AIS receivers who wish to share their data openly or commercially while maintaining anonymity from end users. Counterintuitively, the possibility of anonymously operating AIS receivers  may be most limited in dense urban environments. This impact on data privacy could permanently constrain the availability of open-access and community AIS data in the most critical and relevant regions unless methods can be developed to restore privacy without compromising data quality.

\vspace{6pt} 


\funding{This research received no external funding.}

\dataavailability{The original data presented in the study and research outcomes are openly available at \href{https://dx.doi.org/10.6084/m9.figshare.29037401}{https://dx.doi.org/10.6084/m9.figshare.29037401}. }

\acknowledgments{In this study, we used the {GeographicLib} \citep{Karney2013}, {n-vector} \citep{Gade2010}, {SciPy} \citep{Virtanen2020}, \mbox{{sphstat} \citep{Hachabiboglu2023}}, {scikit-image}, {GDAL}, and {QGIS} software, and the {Basemap}, {Colorcet}, {cmocean}, {Palletable}, and {Font Awesome} packages for visualization. 
We thank \citet{aisstream2025} 
 and the contributing radio operators for providing the AIS data, VesselFinder Ltd. \cite{VesselFinder2025} for granting permission to use their gross tonnage data, and the City of Kawasaki for the kind permission to reproduce their map data. The author wishes to gratefully acknowledge the helpful comments provided by Michael Dziomba, Henrik Hanssen, Marcel Strzys, Hiroki Tsuda, Thomas Zwetyenga, and the anonymous referees.
} 

\conflictsofinterest{The author declares no conflicts of interest. Moritz Hütten is employed by GRID Inc. The author declares that the research was conducted in the absence of any commercial or financial relationships that could be construed as potential conflicts of interest. GRID Inc. approved this manuscript for publication; however, it had no role in the design of the study; in the collection, analysis, or interpretation of data; in the writing of the manuscript; or in the decision to publish the results.
} 


\newpage
\abbreviations{Abbreviations}{
The following abbreviations are used in this manuscript:
\\

\noindent 
\begin{tabular}{@{}ll}
AIS & Automatic Identification System\\
CL & Confidence Level \\
CR & Containment Range \\
GT & Gross Tonnage\\
IMO & International Maritime Organization\\
JAMS & Japan Association of Marine Safety\\
JCG & Japan Coast Guard\\
LNG & Liquefied Natural Gas\\
MLIT & Ministry of Land, Infrastructure, Transport, and Tourism of Japan\\
MMSI & Maritime Mobile Service Identity\\
ROI & Region of Interest\\
SAR & Synthetic-Aperture Radar\\
SISECA & Association for the Environmental Conservation of the Seto Inland Sea\\
UTC & Coordinated Universal Time\\

\end{tabular}
}

\appendixtitles{yes} 
\appendixstart
\appendix

\section[\appendixname~\thesection]{Analysis Details and Parameters}
\label{app:data_processing}

\subsection{AIS Message Cleaning and Classification 
}
\label{app:data_processing1}
\begin{figure}[ht!] 

	\includegraphics[width=\textwidth, trim=0px 0px 0px 15px, clip]{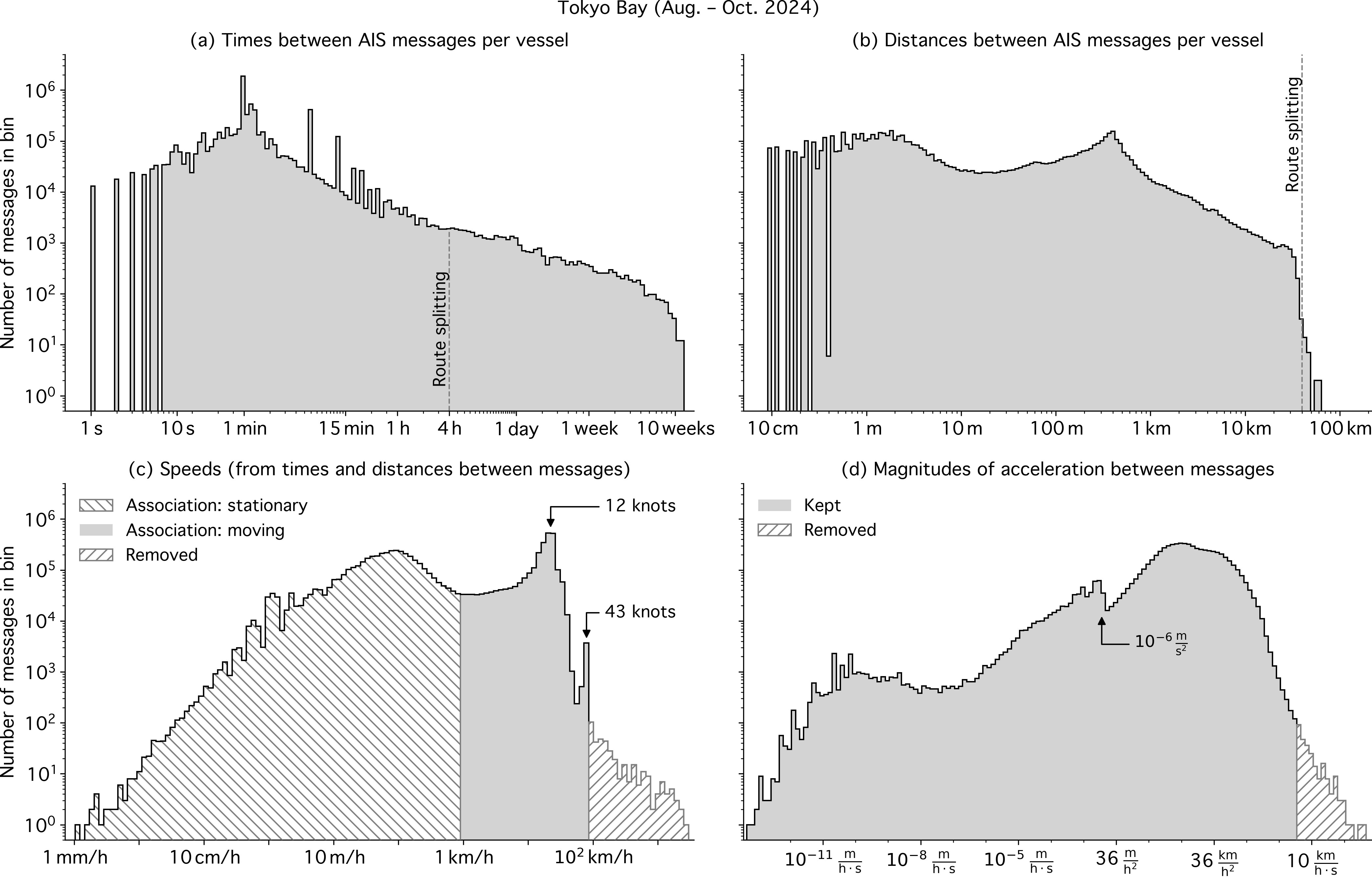} 
	  \caption{
Histograms (with logarithmic binning) of messages, after removal of static vessels and low-speed duplicate messages, showing time differences between consecutive messages (\textbf{a}), distances (\textbf{b}), speeds (\textbf{c}), and accelerations (\textbf{d}).
}
    \label{fig:ais_signals_histograms_tokyo} 
\end{figure}

In this analysis, we removed all static vessels staying within a square of \SI{100}{\meter} side length, the order of magnitude of a vessel's length. The fraction of static-report messages was 15\%, and their position information was adjusted. We removed 0.7\% of low-speed duplicate messages and split the messages into stationary and movement periods when, between two messages, (i) the speed was lower than \SI{0.5}{\knot} \citep{Cerdeiro2020,Li2022,Huetten2025a}, (ii) the time gap was larger than \SI{4}{\hour} (illustrated by the dashed vertical line in Figure$\,$\ref{fig:ais_signals_histograms_tokyo}a), chosen as an upper limit to cross Tokyo Bay on a straight path, (iii) the distance was greater than \SI{40}{\km} (illustrated by the dashed vertical line in Figure~\ref{fig:ais_signals_histograms_tokyo}b), chosen as the corresponding distance upper limit, or (iv) the vessel track entered and again left the gray-hatched transit area shown in Figure~\ref{fig:roi_data_map_tokyo_both}a.
This 
 splitting assigned 38\% of the messages to 100,250 distinct movements; 61\% of messages were classified as stationary, and 0.5\% were discarded as movements entirely within the transit area shown in Figure~\ref{fig:roi_data_map_tokyo_both}a. 

We then iteratively removed all messages where the vessel speed exceeded \SI{50}{\knot} relative to the previous message, and all messages with an associated magnitude of acceleration greater than  \SI{1}{\meter/\square\second}, choosing the same values as in \citep{Huetten2025a}. This removed another 0.1\% of the original messages. Afterward, a second pass of the splitting algorithm and area filter was applied, reducing the number of movement periods by 10\% and reassigning 0.3\% of the messages from moving to stationary periods. Finally, movement periods separated by \SI{120}{\second} or less (twice the most common broadcasting interval) were merged, affecting 16\% of remaining movements and removing an additional 0.2\% of messages. As a result, 38\% of all messages in the ROI were associated with 76,098~movements of 3271 unique vessels; 61\% were associated with stationary periods of those vessels, and a total of 1.5\% of messages were discarded. 

The cleaning and splitting steps are illustrated in Figure \ref{fig:ais_signals_histograms_tokyo}. Figure~\ref{fig:ais_signals_histograms_tokyo}a,b show the splitting conditions, with discrete time and distance values at the lower ends of the histograms reflecting the limited temporal and spatial numeric precision. The spikes in Figure~\ref{fig:ais_signals_histograms_tokyo}a indicate the most common broadcasting interval (\SI{1}{\minute}) and multiples of the 6-minute static-report interval. The second speed peak at \SI{43}{\knot} in Figure~\ref{fig:ais_signals_histograms_tokyo}c corresponds to hydrofoil ferries operating between Tokyo and the Izu Islands. 

The edge in Figure~\ref{fig:ais_signals_histograms_tokyo}d at $\sim\,$$10^{-6}\,$\SI{}{\meter/\square\second} is due to the limited distance resolution and the discrete static-report interval multiples: 
The absolute acceleration $|a| = |\Delta v|/\Delta t$ attributed to a message was calculated with
\begin{align}
 |\Delta v| = |\Delta d_2/\Delta t_2 - \Delta d_1/\Delta t_1|
 \label{eq:spike1}
 \end{align}
 and
 \begin{align}
\Delta t = (\Delta t_1 + \Delta t_2)/2
\label{eq:spike2}
 \end{align}
where $\Delta d_i,\, \Delta t_i$ are the distance and time differences relative to the previous ($i=1$) and next ($i=2$) messages. The chosen six-digit-precision rounding of coordinates resulted, for stationary vessels, in an overabundance of cases with $\Delta d_1\approx\Delta d_2 = \Delta d \approx 10~\text{cm}$. Additionally, from stationary vessels, mostly only static-report messages were received at intervals of $n\times{6}$ min. The multiple $n=1,2,3,\ldots$ indicates that occasionally, a six-minute-interval message was missed or not received. By setting  $\Delta t_1={6}$ min and $\Delta t_2 = n\,\Delta t_1$, one obtained, after combining Equation~(\ref{eq:spike1}), Equation~(\ref{eq:spike2}), 
 and the values for $\Delta d_i,\, \Delta t_i$,
\begin{align}
|a| = \left|\frac{1-n}{n\,(1+n)}\right|\times\frac{\Delta d}{10\,\text{cm}}\times1.6\times10^{-6}\,\SI{}{\meter/\square\second}\,.
 \end{align}

The 
 likelihood of a given $n$ decreases with increasing $n$ (Figure \ref{fig:ais_signals_histograms_tokyo}a), such that the most likely non-zero acceleration becomes, for $n=2$, $|a| =\frac{\Delta d}{10\,\text{cm}}\times2.6\times$$10^{-7}\,$\SI{}{\meter/\square\second}.

\subsection{Trajectory Reconstruction}

\begin{figure}[ht!] 
		\includegraphics[width=0.5\textwidth, trim=0px 0px 0px 15px,clip]{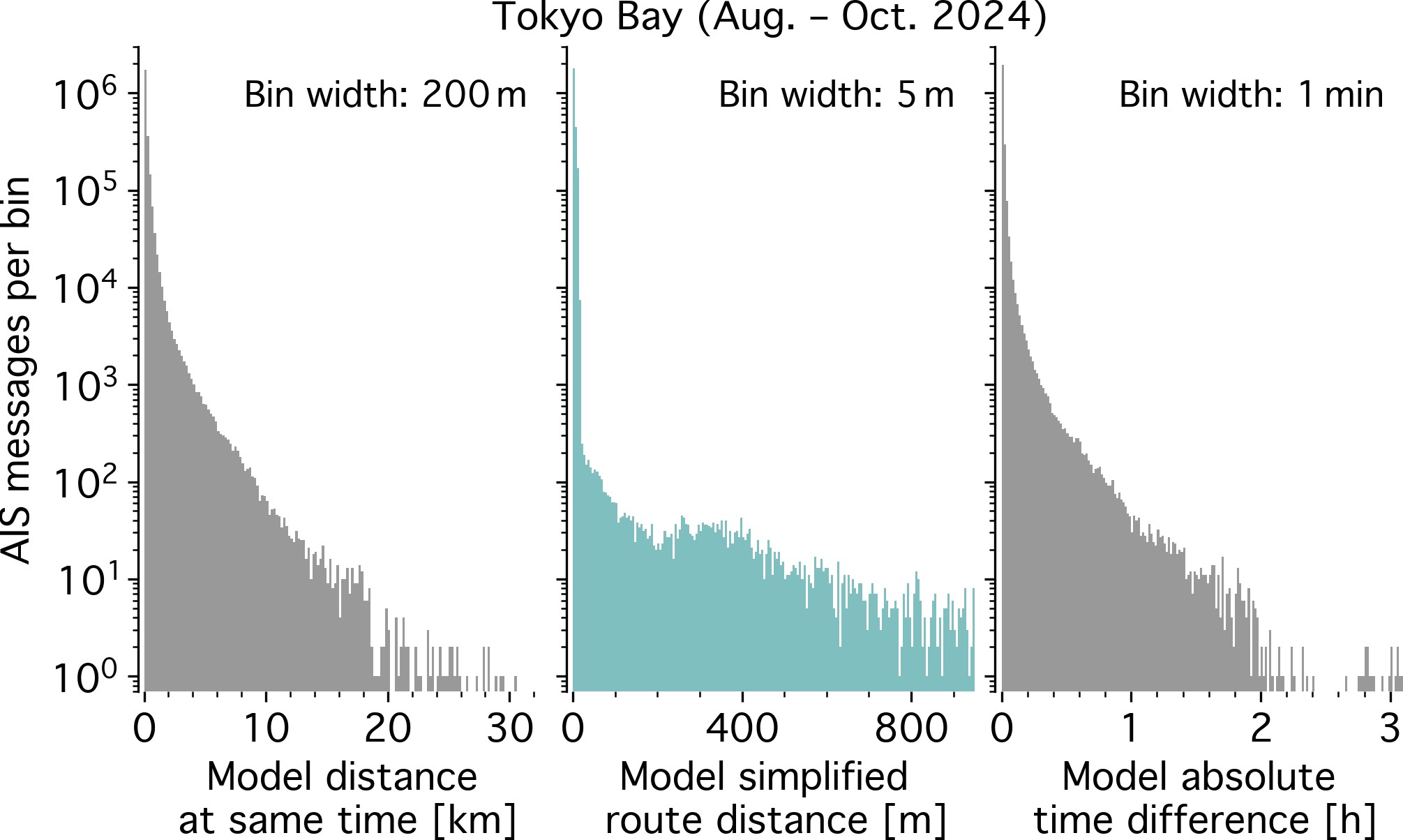} 	  \caption{Comparison of original AIS message positions and times with the trajectory model for moving vessels. \textbf{Left panel}:  Distances between message positions and the vessel model at the message time. \textbf{Middle panel}: Distances between message positions and simplified routes. \textbf{Right panel}: Time differences between message times and the times at which the trajectory model is closest to the message positions. 
      }
      \label{fig:model_analysis_tokyo}
\end{figure}

Movement representations (trajectories) from cleaned messages classified as being broadcast by moving vessels were constructed as described in \cite{Huetten2025a}: Trajectories are represented by geometric tracks (routes) and a speed parametrization along the routes. We adopted a route-simplification tolerance $d_\text{tol} = \SI{10}{\meter}$ (approximately an average vessel width) and, as in \cite{Huetten2025a}, a speed accuracy of 5\%. The trajectory model proves robust across scales, maintaining $\sim1\%$ accuracy in this case with an ROI diameter---and a correspondingly scaled $d_\text{tol}$---reduced by an order of magnitude compared to \cite{Huetten2025a}. This is illustrated by the distributions in Figure \ref{fig:model_analysis_tokyo}, comparing the distance and timing information between the original messages during vessel movements and the modeled trajectories. The left panel of Figure~\ref{fig:model_analysis_tokyo} shows that the modeled vessel positions deviate from the message positions at the same time by a median of \SI{83}{\meter}, with 90\% of deviations below \SI{521}{\meter}. The middle panel shows the deviation without considering timing, i.e., the effective precision of the route simplification, with a median deviation of \SI{1.4}{\meter} and a 90th percentile of \SI{9.0}{\meter}. The right panel illustrates the model's timing precision: for each AIS message, the time at which the model trajectory comes closest in space differs by \SI{21}{\second} (median) and \SI{102}{\second} (90th percentile). Relative to the total trajectory distances and durations, the medians of the deviations in the right and left panels are both 0.7\%, thus below 1\%. 

\begin{figure}[ht!]
    \includegraphics[width=0.5\textwidth, trim=0px 0px 0px 15px,clip]{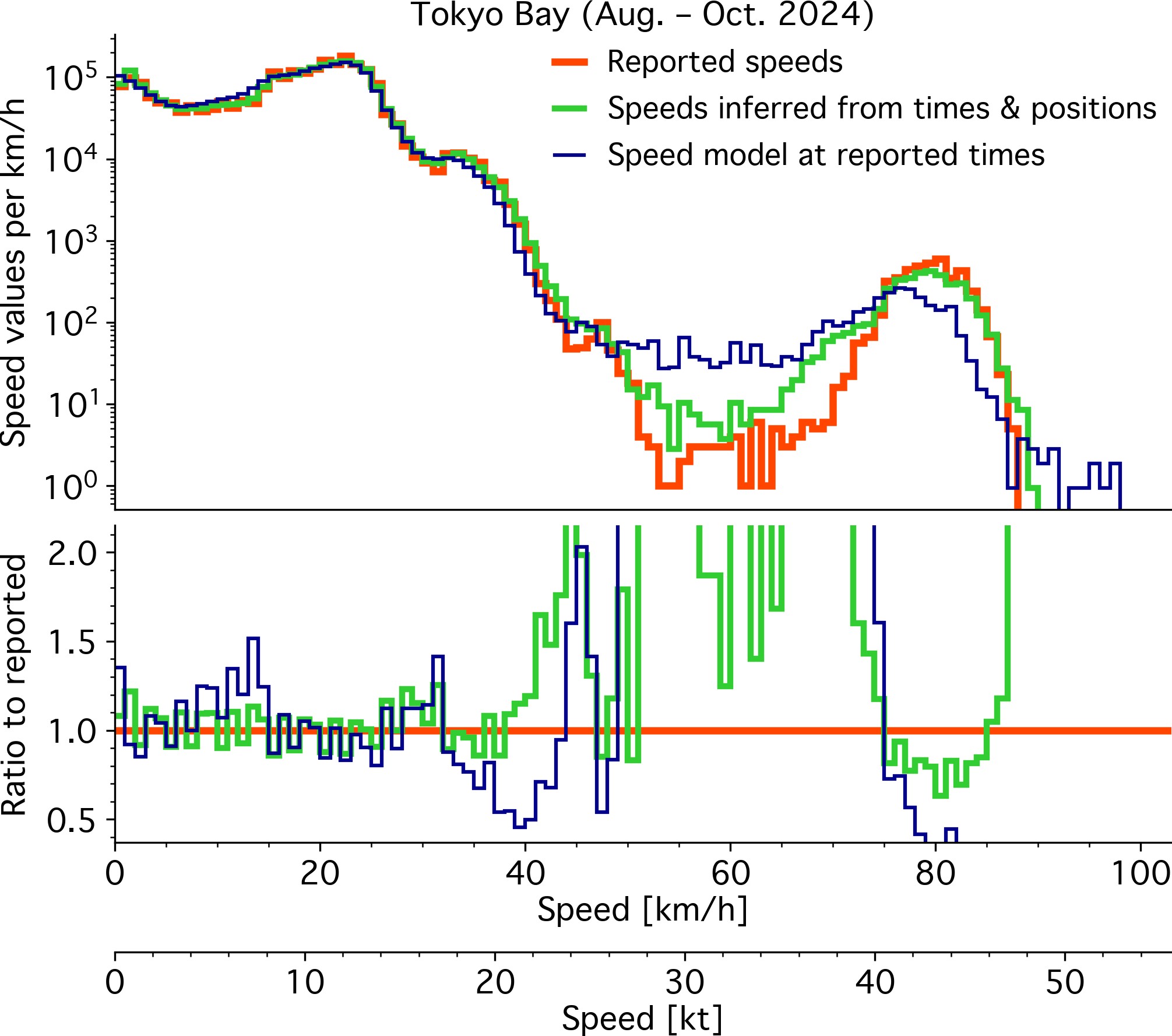}     \caption{
          Vessel speeds from the AIS position-report messages (thick red curves),  compared with speeds calculated from reported positions and times (green curves) and with the trajectory model at the message times (thin blue curves), shown absolute (top panel) and relative (bottom panel).     } 
    \label{fig:speed_analysis_tokyo}
\end{figure}

Figure \ref{fig:speed_analysis_tokyo} additionally compares the speed information between the original messages and the modeled trajectories. The red thick curve shows the speeds reported by the vessels, the green curve shows speeds inferred from position and timing information (same data as in Figure~\ref{fig:ais_signals_histograms_tokyo} without static-report messages and binned in linear intervals), and the blue thin curve shows the model speeds at the corresponding message times. As in \cite{Huetten2025a}, the model accurately captures vessel motion at speeds $\lesssim$40 km/s but deviates from observed values at the top 0.3\% of speeds in the Tokyo Bay dataset.

\subsection{Computation of Metrics}

We also computed vessel metrics (momentary vessel statistics, transits, and two-dimensional grids), analogous to \cite{Huetten2025a}. A total of 90\% of the observed vessels were registered with an International Maritime Organization (IMO) number (``IMO vessels''), and for 99\% of these, we obtained the gross tonnage from VesselFinder Ltd.~\cite{VesselFinder2025}. We excluded the first and last three days of the period from the all-vessel and stationary average calculations (hatched areas in Figure~\ref{fig:ais_signals_time_tokyo}) to account for temporal edge effects caused by the limited analysis time window. For the gridded maps, we sampled the vessel trajectories at intervals of \SI{10}{\meter}. We used the shorelines from \citet{Openstreetmap2024}, rasterized at a resolution of \SI{1}{\meter}, to exclude vessel tracks over land, and we discarded stationary vessel periods from the map representation if the distance between the enclosing leg end and start coordinates was larger than \SI{500}{\meter}. As displayed in Figure~\ref{fig:roi_data_map_tokyo_both}b, this corresponds to the diameter scale on which anchored vessels may drift during mooring periods with changing wind direction.

\subsection{Berth Identification}

We applied the density-thresholding algorithm using the same parameters as in \cite{Huetten2025a}. We denoised the density map by Gaussian smoothing over 1.5 cells and applied a watershed algorithm \citep{scikit-image} to extended regions above half the seed density threshold to find the corresponding overdensity areas. 
We then searched for seed locations above a density threshold of 10 vessels/\SI{}{\square\km}, which is 20 times the value used for port detection in \cite{Huetten2025a} and corresponds approximately to the difference in spatial scale between the two analyses. For these defined overdensity areas, we counted arrivals by the number of trajectory endpoints falling within an area and assigned the most likely berth or anchorage name and vessel type based on the arriving vessels' destination and type information. We classified all areas with centers of mass within \SI{200}{\meter} of the shore as berth locations.

\section[\appendixname~\thesection]{Supplementary Tables and Figure}
\label{app:receiver_loc}

\vspace{-6pt}
\begin{table}[ht!]
\caption{
Average number of vessels in the ROI ($N$) and total inbound and outbound transit rate, $\dot{N}$, for the default (\textit{df}) model, and \textit{low}/\textit{hi} systematic uncertainty estimates. The last two rows show the relative difference from the \textit{df} case.
}\label{tab:case_comparison}
\begin{tabularx}{\textwidth}{CCC}
\toprule
 & \textbf{Vessels in ROI, $\boldsymbol{N}$} & \textbf{Transits/Day, $\boldsymbol{\dot{N}}$} \\
\midrule
Case \textit{low} & $370.8$ & $312.9$ \\
Default (case \textit{df})   & $381.0$ & $292.7$ \\
Case \textit{hi}   & $391.6$ & $282.4$ \\\midrule
$\delta_\text{low}$ & $-2.7\%$ & $+6.9\%$ \\
$\delta_\text{hi}$  & $+2.8\%$ & $-3.5\%$ \\
\bottomrule
\end{tabularx}
\end{table}\vspace{-9pt}

\begin{table}[ht!]
\caption{Fraction of AIS-B vessels, $\delta_\text{ais-b}$, among all vessels operating in Tokyo Bay, estimated from the October 2024 Tokyo Bay AIS data.}
\label{tab:vessel_aisb}
\begin{tabular*}{\textwidth}{lr}
\toprule 
\textbf{Vessel Type} & \boldmath$\boldsymbol{\delta_\text{\textbf{ais-b}}}$
  \\
\midrule
 All vessels & 
  $12\%\;$  
  \\\midrule
 Passenger, high-speed & 
 $0\%$
 \\[0.15cm]
 Law enforcement, military & 
 $0\%$ 
 \\[0.15cm]
 Cargo  & 
 $5\%$ 
 \\[0.15cm]
 Pilot, tug, rescue, diving/dredging  & 
 $20\%$
  \\[0.15cm]
 Tanker  & 
 $7\%$ 
 \\[0.15cm]
 Others, including fishing (derived) & 
 $37\%\;$
 \\\midrule
 GT $<10{,}000$ (derived) & $14\%$ \\[0.15cm]
 GT $\geq10{,}000$ (assumed) & $0\%\;$
 \\
\bottomrule
\end{tabular*}
\end{table}\vspace{-9pt}

\begin{table}[ht!]
\caption{Reference and inferred AIS receiver station positions. $\varphi$ and $\lambda$ denote latitude and longitude coordinates. $a_{95\%}$, $b_{95\%}$, and $\theta$ denote the major and minor semi-axes and local orientation (clockwise from true north) of the mean and median confidence ellipses, respectively. This corresponds to$(a_{95\%},\,b_{95\%})=\sqrt{{(m\,F_{m,\,n-m,\,0.95}})/{(n-m)}}\;(a,\,b)$ according to Equation~(\ref{eq:confidence_ellipse}). $^\dagger$ The Tokyo receiver has been offline since 29 August 2025 \cite{AISHub2025}. $^\ddagger$ JR1CAD provides its position at meter precision; however, this information is omitted here for privacy reasons. $^\S$ No confidence ellipse could be calculated due to low statistics.
}
\label{tab:receiver_positions}
\begin{adjustwidth}{-1.5cm}{0cm}
\newcolumntype{C}{>{\centering\arraybackslash}X}
\begin{tabularx}{\fulllength}{llCCCCCl}
\toprule 
\textbf{Station} &
\textbf{Position Name} & 
$\boldsymbol{\varphi}$ \textbf{[}$\boldsymbol{^\circ}$\textbf{]} & 
$\boldsymbol{\lambda}$ \textbf{[}$\boldsymbol{^\circ}$\textbf{]} &  
$\boldsymbol{a_{95\%}}$  &  
$\boldsymbol{b_{95\%}}$ &  
$\boldsymbol{\theta}$ \textbf{[}$\boldsymbol{^\circ}$\textbf{]} & 
\textbf{Source}
  \\\midrule
  Tokyo  & AISHub reported position $^\dagger$ & $35.614\;\,$ & $139.631\;\,$ & $\SI{550}{\meter}$ & $\SI{550}{\meter}$ & $/$ & \cite{AISHub2025} \\
 & Futakotamagawa  Rise Tower East & $35.6089$ & $139.6313$ & $\;\,\SI{20}{\meter}$ & $\;\,\SI{20}{\meter}$ & $/$ & \cite{TallBuildings2025} \\
 & Weighted mean focal point  & $35.6106$ & $139.6327$ & $\SI{270}{\meter}$ & $\;\,\SI{70}{\meter}$ & $-25.9$ &  this analysis \\
  & Weighted median focal point  & $35.6116$ & $139.6324$ & $\SI{149}{\meter}$ & $\;\,\SI{58}{\meter}$ & $-31.7$ &  this analysis \\
 & Unweighted mean focal point  & $35.6097$ & $139.6313$ & $\SI{13.60}{\km}$ & $\SI{1.37}{\km}$ & $-21.6$ &  this analysis \\
 & Unweighted median focal point  & $35.6112$ & $139.6325$ & $\SI{227}{\meter}$ & $\;\,\SI{71}{\meter}$ & $-41.2$ &  this analysis \\\midrule
  Yokohama  & VesselFinder reported position $^\ddagger$ & $35.475\;\,$ & $139.549\;\,$ & $\SI{550}{\meter}$ & $\SI{550}{\meter}$ & $/$ & \cite{VesselFinder2025} \\
 & Weighted mean focal point  & $35.4708$ & $139.5876$ & $\SI{10.71}{\km}$ & $\SI{133}{\meter}$ & $-70.7$ &  this analysis \\
  & Weighted median focal point $^\S$  & $35.4661$ & $139.5997$ & $/$ & $/$ & $/$ &  this analysis \\
 & Unweighted mean focal point  & $35.4912$ & $139.5292$ & $\SI{33.36}{\km}$ & $\SI{201}{\meter}$ & $-70.7$ &  this analysis \\
 & Unweighted median focal point $^\S$  & $35.4670$ & $139.5971$ & $/$ & $/$ & $/$ &  this analysis \\\bottomrule
\end{tabularx}
\end{adjustwidth}
\end{table}

\begin{figure}[ht!]
    \includegraphics[width=0.5\textwidth]{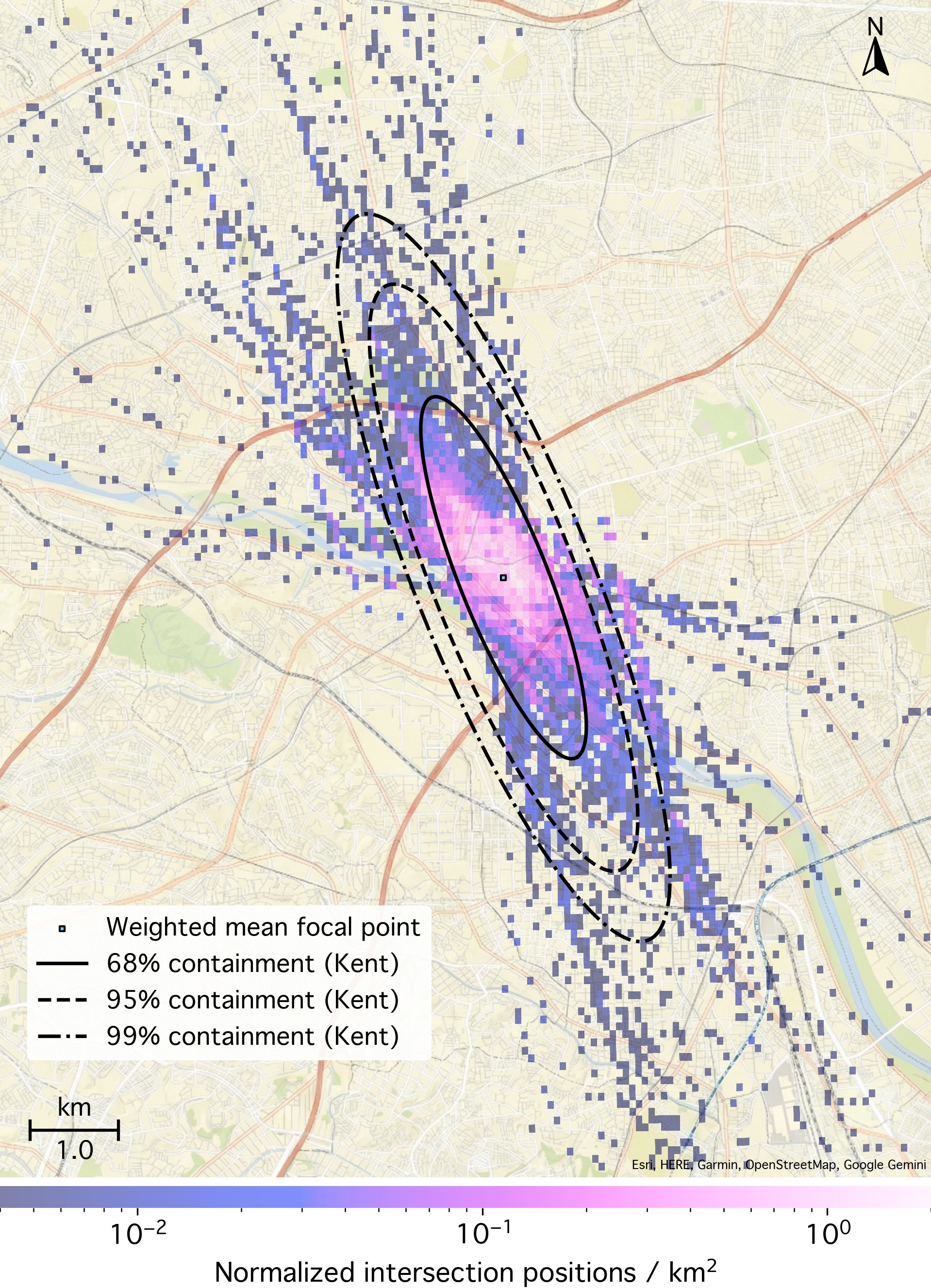}
    \caption{Pairwise 
 intersections of 190  geodesic sections fitted to the AIS  message shadows associated with the Tokyo receiver, weighted by intersection angles. Black curves show the containment ranges of a Kent distribution fitted to the intersection positions. Most intersections (99\%) fall within the figure, indicating wider tails than normal-distributed values (modeled by a Kent distribution), for which 99\% would lie within the dashed-dotted ellipse. The figure covers \SI{100}{\square\km} around the weighted mean position (Table~\ref{tab:receiver_positions}) in the Tokyo Metropolitan Area.    } 
    \label{fig:mysterious_receiver_wide_weighted}
\end{figure}

\newpage

\isPreprints{}{
\begin{adjustwidth}{-\extralength}{0cm}
} 
\reftitle{References}

\isPreprints{}{
\end{adjustwidth}
} 
\end{document}